\documentclass{aa}

\def\esn{\dot \epsilon_{\rm SN}}
\def\flx{{\cal F}}

\def\mb{M_{\rm b}}
\def\mc{m_{\rm c}}
\def\mde{\dot M_{ev}}

\def\msun{M_{\odot}}
\def\nc{N_{\rm c}}

\def\rat{{\cal R}}
\def\nsn{{\cal N}}

\def\rc{R_{\rm c}}

\def\rcpc{r_{\rm c,pc}}
\def\rhocl{\dot \rho_{\rm cl}}
\def\rhosn{\dot \rho_{\rm SN}}

\def\vc{v_{\rm c}}
\def\v2c{v_{\rm c}^2}
\begin{document}
\input{epsf}
\newbox\grsign \setbox\grsign=\hbox{$>$} \newdimen\grdimen \grdimen=\ht\grsign
\newbox\simlessbox \newbox\simgreatbox
\setbox\simgreatbox=\hbox{\raise.5ex\hbox{$>$}\llap
     {\lower.5ex\hbox{$\sim$}}}\ht1=\grdimen\dp1=0pt
\setbox\simlessbox=\hbox{\raise.5ex\hbox{$<$}\llap
     {\lower.5ex\hbox{$\sim$}}}\ht2=\grdimen\dp2=0pt
\def\simgreat{\mathrel{\copy\simgreatbox}}
\def\simless{\mathrel{\copy\simlessbox}}
\newbox\simppropto
\setbox\simppropto=\hbox{\raise.5ex\hbox{$\sim$}\llap
     {\lower.5ex\hbox{$\propto$}}}\ht2=\grdimen\dp2=0pt
\def\simpropto{\mathrel{\copy\simppropto}}
\title{Evolution of the ISM of Starburst galaxies: the SN heating efficiency}
\author{
Melioli C.\inst{1}
\and
de Gouveia Dal Pino E.M.\inst{1}
}
\offprints{Melioli C.}
\institute{
Universidade de S\~ao Paulo, IAG, Rua do Mat\~ao 1226,
Cidade Universit\'aria, S\~ao Paulo 05508-900, Brazil\\
 e-mail: cmelioli@astro.iag.usp.br , \ \ dalpino@astro.iag.usp.br
}

\date{Received ; accepted}

\abstract{The interstellar medium heated by supernova explosions (SN)
may acquire an expansion velocity larger than the escape velocity and
leave the galaxy through a supersonic wind. Galactic winds are
effectively observed in many local starburst galaxies. SN ejecta are
transported out of the galaxies by such winds which thus affect the
chemical evolution of the galaxies.  The effectiveness of the
processes mentioned above depends on the heating efficiency (HE) of
the SNe, i.e. on the fraction of SN energy which is not radiated
away. The value of HE, in particular in starburst (SB) galaxies, is a
matter of debate.  We have constructed a simple semi-analytic model,
considering the essential ingredients of a SB environment which is
able to qualitatively trace the thermalisation history of the ISM in a
SB region and determine the HE evolution.  Our study has been also
accompanied by fully 3-D radiative cooling, hydrodynamical simulations
of SNR-SNR and SNR-clouds interactions.  We find that, as long as the
typical time scale of mass-loss of the clouds to the ambient medium, which is
often dominated by photoevaporation, remains shorter than the time scale at
which the
SNRs interact to form a superbubble, the SN heating efficiency remains very
small, as radiative cooling of the gas dominates.
If there is a continuous production of clouds by the
gas swept by the SNR shells,
this occurs during
the first $\leq $16 Myrs of the SB activity (of $\sim$ 30 Myrs), after
which the efficiency rapidly increases to one, leading to a possible
galactic wind formation.
Under an extreme condition in which no clouds are allowed to form,
other than those that were already initially present in the  SB environment,
then in this case HE increases to one in only few Myrs.
We conclude that the HE value has a
time-dependent trend that is sensitive to the initial conditions of
the system and cannot be simply assumed to be $\sim$ 1, as it is
commonly done in most SB galactic wind models.

}

\titlerunning{SN heating efficiency of the ISM of SB}
\authorrunning{Melioli et al.}
\maketitle

\section{Introduction}

Energization of the interstellar medium (ISM) by supernova (SN)
explosions covers a crucial role in a number of astrophysical
situations. During the galaxy formation, the first exploding stars may
heat and expand the local gas reducing, or even halting the ongoing
star formation. Galaxies with highly active star formation regions,
starburst (SB) galaxies, present a high SN rate. The gas heated by SNe
may acquire an expansion velocity larger than the escape velocity
and leave the galaxy through a supersonic wind. Galactic winds are
effectively observed in many local SB galaxies (Lehnert \&
Heckman 1996). SN ejecta are transported out of the galaxies by
such winds which thus affect the chemical evolution of the galaxies.

The effectiveness of the processes mentioned above depends on the
heating efficiency (HE) of the SNe, i.e. on the fraction of the
SN explosion energy that remains effectively stored in the ISM gas and
is not radiated away.
In a SB region, several SN explosions occur at a high rate
inside a relatively small volume, and their remnants are likely to
interact with each other. Larson (1974), in his pioneering work
of the monolithic galaxy formation, compares the cooling time, $t_{\rm
c}$, defined as the age of a SNR when 50 per cent of its thermal
energy has been radiated away, to the SNRs interaction time, $t_{int}$, defined
as the time it takes for the remnants to occupy $\sim$ 60 per cent of
the volume of the SB region.
After an interaction time, the remnants have
collided with each other and their evolution ceases.  Radiative losses
are important if $t_{\rm c}<t_{int}$. Larson evaluates that, in
the cases of interest, the energy which is effectively transmitted
to the gas is $\sim 10$ per cent of the explosion energy. This value
of the HE has been adopted in a number of SB models (e.g. Babul
\& Rees 1992, Murakami \& Babul 1999).  Bradamante, Matteucci \&
D'Ercole (1998), following a different line to estimate HE, have considered
only the final fraction of the SN energy that effectively turns into thermal
energy of the ambient medium when the remnant, already at the end of its
radiative phase, stalls. This gives HE $\sim$ 0.03 in the cases of
interest, and this value has been adopted by Ferrara \& Tolstoy (2000) and
by Recchi, Matteucci and D'Ercole (2001) in their models of dwarf
galaxies.

These low values of the HE have been questioned by several
authors. Following Cox \& Smith (1974), Larson (1974) warned that, for a high
SN rate, the collisions of the first generation of SNRs possibly lead to
the formation of an interconnected network of tunnels containing
very hot, tenuous gas maintained at high temperature by continuing
supernova explosions. The successive generations of SNRs are likely to
produce a value of the HE close to unity. In fact, the remnants do not
reach high densities during their expansion, and their radiative
losses are expected to remain negligible as the emissivity is proportional
to
the square of the density.
Dekel \& Silk (1986) find that the SNRs interact
before entering in the radiative phase, and almost all their energy is
transferred to the ISM gas. Short interacting times seem to invalidate the
assumption of stalling remnants by Bradamante, Matteucci \&
D'Ercole (1998). Strickland \&
Stevens (2000) argue that with a value of the HE as low as that
proposed by Bradamante, Matteucci \& D'Ercole (1998) the galactic wind
observed in M82 (the best studied SB with a galactic wind) could be hardly
driven. Chevalier \& Clegg (1985) find that 100 per cent is a likely
value for the HE in M82.

In the absence of a clear response from observations (e.g. Della Ceca et al.
1999; Cappi et al. 1999), the above considerations have led to the
widespread
belief that the HE within a SB region must be quite high to sustain galactic
winds. For this reason, most of the simulations of galactic winds
found in literature assume a HE value of 100 per cent (e.g. Suchkov et
al. 1994, Silich \& Tenorio-Tagle 1998, D'Ercole \& Brighenti 1999,
Mac Low \& Ferrara 1999, de Gouveia Dal Pino \& Medina Tanco 1999,
Strickland \& Stevens 2000).
However, several of these ``standard'' models fail in reproducing the sizes
of
the active regions, which turn out to be larger than those observed
(see, e.g.
Tenorio-Tagle \& Mu\~noz-Tu\~n\'on 1998 and references therein).
Moreover, Recchi, Matteucci \& D'Ercole (2001) (also Tosi 2003), 
are able to
reproduce the chemical and dynamical characteristics of the SB IZw18, for
example,
only assuming, in an instantaneous SB, a HE value of 0.03,
which is about 30 times lower  than
the value commonly assumed in the simulations of SB galactic winds.
In models with HE=1, all the gas in IZw18 is lost, in opposition to what is
observed.
We are thus motivated to investigate further the coupling between SNRs and
the ISM
inside active regions, and explore under which circumstances the HE
may assume low values. In order to do this, we will not be concerned
with a detailed description of the structure of the star-bursting galactic
region.
Instead, we construct a model, as simple as possible, disregarding all
the details that are not essential for our purposes; such a model is highly
idealized, but still able to give us insights on the thermalisation history
of the ISM inside a star-bursting region.

In the following sections, we outline the characteristics of the model and
present solutions for a steady state system ($\S$ 2);
discuss the physical processes in a SB environment including the
 presentation
of the relevant results of 3-D numerical simulations of the interactions
between SNRs and clouds ($\S$ 3); and study the relevant time scales for the
mixing  of the several gaseous components of the system ($\S$ 4).
In $\S$ 5, we present the results of our evolution model for a SB
environment including all the physical processes described in the previous
sections; and finally in $\S$ 6, we present a brief discussion and draw
our conclusions.

\section{The model}
\subsection{Assumptions}
The events occurring in a SB region are quite complex. Supernova
remnants (SNRs) expand through a diffuse medium in which clouds are
embedded. These clouds suffer mass loss by ablation, thermal
evaporation and photoevaporation due to the huge amount of ionizing
photons emitted by massive stars and by SNs.  The gas is heated by
such SNs, and is cooled via radiative losses. These losses depend on
the square of the gas density, so that two competing facts occur: on
one hand the SNRs sweep the diffuse gas, thus lowering its mean
density in the SB region; on the other hand the gas density is
increased by mass loss from the clouds. The importance of the radiative
cooling depends on which of these two processes dominate. If the cloud
mass loss rate is smaller than the sweeping rate by SNRs, the
successive SNRs expand in a medium more and more rarefied, thus
remaining adiabatic and eventually transferring most of their energy
to the ISM; in this case the HE would be close to unity. On the
contrary, HE would be very low for a cloud mass loss rate larger than the
sweeping rate.

It is very difficult to make numerical simulations of the above
scenario taking into account all the physical processes. Moreover, the
length scales involved are quite different, ranging from hundredths of
parsec of the conduction front width at the edge of the clouds (see
next section) to the size of the SB region, typically 100 pc
(e.g. Meurer et al. 1995, Planesas, Colina \& Perez-Olea 1997);
numerical simulations
able to cover this range of length scales with appropriate spatial
resolution would require an unrealistic amount of cpu time. We thus
follow a semi-analytical approach, as simple as possible, but
incorporating the ``micro" physics as found in literature.
We refer to previous works that have addressed the evolution of a 
multiphase ISM assuming realistic cooling and heating processes and a number 
of mechanisms for clouds destruction and formation (see, e.g., Bertoldi \& 
McKee 1990, Rosen \& Bregman 1995; V\'azquez-Semadeni, Pasot \& Pouquet 1995; 
Shore \& Ferrini 1995; Wada \& Norman 1999, Wada \& Norman 2001), though they 
were mostly concerned with the ISM of our Galaxy and have focused on regions 
with typical sizes and physical conditions which are generally distinct from 
those addressed here, most of these studies have provided the guide lines for 
building the present approach.

We consider the following toy model. An instantaneous burst of star
formation occurs inside a spherical region
of radius $R_{\rm SB}$. Assuming the Salpeter initial mass function
(IMF) the number $\nsn$ of stars with mass greater than $8 \msun$
(i.e., the SN number) is $\nsn \sim 0.01(\mb/\msun)$, where $\mb$ is
the total mass of the stars in the SB. The type II supernova
(SNII) activity lasts up to a time $t_{\rm b} \sim 3\times 10^7$ yr, which
is the lifetime of an 8 $\msun$ star. A constant SN rate is thus given by
$\rat = \nsn/t_{\rm b}$, which is in good agreement with more accurate
evaluations given by Leitherer et al. (1999).
It follows that the mass deposition rate and the energy
injection rate are given, respectively, by $\dot M_{\rm SN}=10\rat$
$\msun$ yr$^{-1}$ and $\dot E_{\rm SN}=10^{51}\rat$ erg yr$^{-1}$,
where we have assumed an energy output of 10$^{51}$ erg and a mass return of
10$\msun$ per SNII.

Assuming a star formation efficiency of $\sim 10$ \% (Colina, Lipari
\&  Macchetto 1991)
and similar characteristics to those observed in the Rosette cloud, where
78 \% of the gas is in the form of compact globules (Williams,
Blitz \& Stark 1995), we obtain that the total mass of clouds in
the SB region is $M_g = 8 \mb$.

The diffuse gas is assumed to be optically thin, and the cooling
function is approximately $\Lambda(T) \cong 1.6\times 10^{-19}\beta
T^{-0.5}$ erg cm$^3$ s$^{-1}$, for $10^5$ K $\leq T \leq 10^{7.5}$ K
(McKee \& Begelman 1990), where the factor $\beta$ is one if the gas
is in ionization equilibrium. At the conductive interfaces,
non-equilibrium ionization effects can increase the cooling up to an
effective value $\beta \sim 10$ (Borkowski, Balbus \& Fristrom 1990,
McKee \& Ostriker 1977).
Following Mathews \& Bregman (1978), in the
range $10^4$ K $\leq T \leq 10^5$ K, we assume a linear dependence for
$\Lambda(T)$ with $T$, $\Lambda(T) = 5.35\times 10^{-27} T$ erg cm$^3$
s$^{-1}$.

As a further simplification, we assume that all the quantities are
uniformly distributed inside the spherical SB volume, and that the gas
leaves this region with a Mach number ${\cal M}=1$, i.e. the gas moves
with a velocity equal to its sound speed (cf. Chevalier \& Clegg 1985),
corresponding to a free expanding steady state wind
 extending to infinity.
A schematic picture of the model is shown in Figure 1 and
Table 1 summarizes the typical values of the parameters of a SB environment.

The aim of the model is to calculate the HE defined as the ratio of
the enthalpy flux through the boundary of the SB region to the SN
luminosity (see eq. (29)).  As a first step, we present in the next
subsection possible steady solutions under very simplistic
assumptions. Despite its excessive idealization, this study outlines
the role of the mass return from the clouds.

\begin{figure}
\centering
\epsfxsize=8cm
\epsfbox{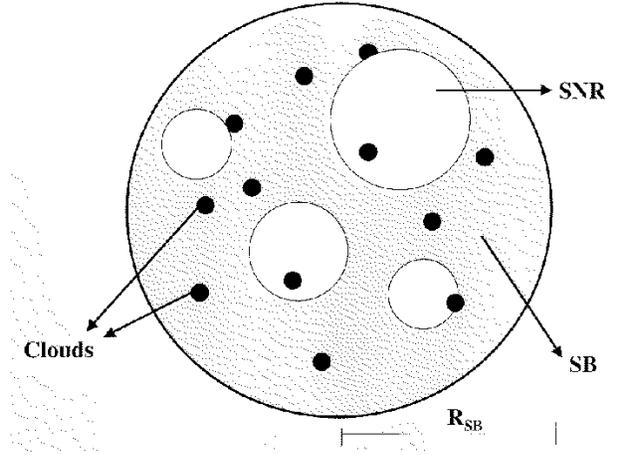}
\caption{Schematic picture of the model: a spherical SB with radius
$R_{\rm SB}$ hosting expanding SNRs and clumps of dense and cold clouds}
\end{figure}

\begin{table*}
\caption[1]{Values for the characteristic parameters of a SB galaxy assumed
in this study}
{\small
\centerline{
\begin{tabular}{|c|c|c|c|c|c|c|}
\hline\hline
\noalign{\smallskip}
\hbox{Model} & \hbox{$R_{SB} \ ({\rm pc})$} & \hbox{$\mb \ (M_{\odot}$)} &
\hbox{$M_g \ (M_{\odot}$)} & \hbox{$\cal N$}
& \hbox{$\cal R \ ({\rm yr^{-1}}$)} & \hbox{$t_b$ (Myr)}\\
\noalign{\smallskip}
\hline
\noalign{\smallskip}
\hbox{Spherical} & \hbox{100-700} & \hbox{$10^5-10^7$} &
\hbox{$10^6-10^8$} &
\hbox{$10^3-10^5$} & \hbox{$3 \times 10^{-5}-3 \times 10^{-3}$} & \hbox
{30}\\
\noalign{\smallskip}
\hline
\noalign{\smallskip}
\end{tabular}
}}
\end{table*}

\subsection{Steady state solutions}

In this simple example, we do not deal with the physics of SNR
expansions and of mass loss from the clouds, but simply consider
SNs and clouds as sources of energy and mass.
Assuming that the gas moves with a velocity equal to its sound
speed, mass and energy conservation then yield:

\begin{equation}
{d\rho\over dt}+{3\over R_{SB}}\rho c_s=\rhosn +\rhocl
\end{equation}
\begin{equation}
2{dp\over dt}+{9\over {\rm R_{SB}}}p c_s=-{\rho^2 \over (\mu m_{\rm H})^2}
\Lambda(T) +{1\over 2}(\rhocl) \vc^2+\esn
\end{equation}
\noindent
where $\rho$ and $p$ are the density and the pressure of the diffuse
ambient gas, $\rhosn$ represents the mass injection rate by SNe,
$\rhocl$ is the mass loss rate by the clouds due to their destruction,
$m_{\rm H}$ is the proton mass, $\mu= 1.3$ is the mean mass per nuclei
of the ionized gas
assumed with 90 \% H and 10 \% He abundances,
$c_s=(2.1 k /\mu m_{\rm H})^{1/2}T^{1/2}$ is the
isothermal sound speed, $R_{SB}$ is the radius in parsec of the SB,
$\esn$ is the rate of energy injection by the SNe explosions. $\vc$ is
the mean cloud velocity; at this stage we do not make any assumption
about the mechanisms responsible for the cloud mass-loss, and simply
assume $\vc \simeq 10^6$ cm s$^{-1}$ as a fiducial value.

In order to obtain possible steady solutions of the equations (1)-(2),
we drop the time derivatives and, after some simple algebra, we
obtain

\begin{equation}
3(1+\xi)c_s^2+{R_{SB}^2 \over 9}\rhosn (1+\xi)^2L(c_s)={1 \over 2}\xi \vc^2+
{\esn \over \rhosn}
\end{equation}
\noindent
where $\xi=\rhocl / \rhosn$, $k$ is the Boltzman constant, and

\begin{equation}
L(c_s)=\cases{{\cal K}{\hat{c_s}^{-3}} & if $T \leq 10^5$ K\cr
     {\cal K}c_s^{-3} & if $T>10^5$ K \cr}
\end{equation}
\noindent
Here ${\cal K}=1.6 \times 10^{-19}(2.1k)^{1/2}(\mu m_{\rm
H})^{-5/2}$ cm$^6$ s$^{-4}$ g$^{-1}$ and $\hat{c_s}$ is the sound speed at
$T=10^5$ K.

The l.h.s. of equation (3) describes the gas cooling. In particular,
the first term represents the cooling due to the adiabatic expansion,
while the second term represents the radiative losses. The
r.h.s. contains the heating sources. The second term represents the SN
heating per unit mass of the ejecta. We note that this term is an
intrinsic property of the SNe and is independent of any assumption
about the SB.  The first term represents the thermalisation of the
kinetic energy of the gas lost by the clouds moving with a mean
velocity $\vc$.  We note that this term remains negligible with
respect to the second one for any reasonable value of $\xi$ and $\vc$.

Figure 2 illustrates the heating sources and the energy sinks of
equation (3) as functions of T for a model with $R_{\rm SB}=100$ pc
and $M_{\rm b}=10^6$ $\msun$. The three solid curves represent the
l.h.s. of equation (3) for three different values of $\xi$ and for
$\beta=1$; the dashed curves hold for the same values of $\xi$, but
for $\beta=10$. The horizontal line represents the r.h.s. of equation
(3) and is virtually independent of $\xi$. At high temperatures,
radiative losses are negligible and the gas cools via adiabatic
expansion (the rising branch of the cooling curve in Fig. 2). The
opposite is true at intermediate temperatures, and the radiative
cooling increases with decreasing temperature. For $T<10^5$ K, cooling
is constant and independent of the temperature.

\begin{figure}
\centering
\epsfxsize=8cm
\epsfbox{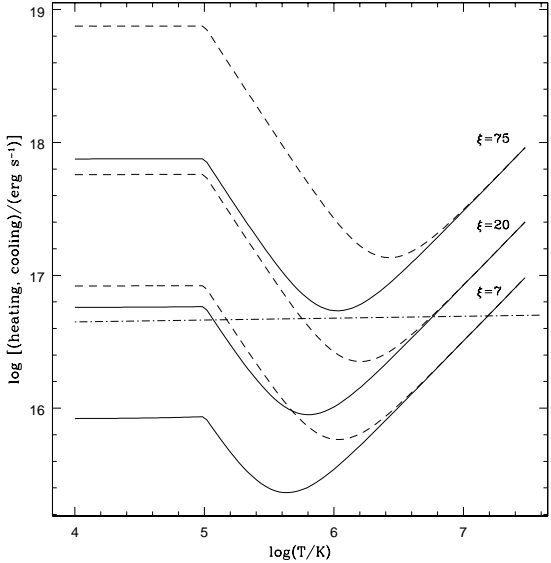}
\caption{Solid lines: cooling curves for $\beta =1$ and three different
values of $\xi$.
Dashed lines: the same for $\beta=10$.
The heating term is represented by the horizontal line which does not depend
on $\beta$ and is virtually independent of $\xi$.}
\end{figure}

In Fig. 2
{\footnote {We notice that due to the uncertainties associated to the
value of
$\beta$ in the radiative cooling curves of Fig. 2 (e.g., Cowie, McKee, \&
Ostriker 1981,  Borkowski, Balbus \& Friston 1990),
we have adopted $\beta \simeq 1$ in all the calculations hereafter, as it
gives a lower limit condition for the gas cooling (see below).}},
 three different cases are recovered, depending on the value
of $\xi$. For $\xi < \xi_{\rm h}$ (see below) only one steady solution
exists (the intersection point between the heating line and the
cooling curve in Fig. 2) and it occurs at a high temperature (the hot
solution).  As pointed out above, at high temperatures radiative
losses are negligible, and the HE value is close to 1. Following Field
(1975), it is easy to show that this solution is stable. For $\xi_{\rm
w} > \xi > \xi_{\rm h}$, two solutions are possible, the hot solution
and the warm solution. This latter solution has very low values of HE
but is highly thermally unstable.  Numerical integration of equations
(1)-(2) shows that the system moves toward higher or lower
temperatures depending on the value of $\xi$ (of course, lower values
of $\xi$, and hence of density, favour a decrease of the emissivity
and an increase of the temperature).  Finally, for $\xi>\xi_{\rm w}$,
no steady solution is possible, and the system moves toward low values
of temperature and HE, for any reasonable choice of initial
conditions.

For the model shown in Fig. 2, we have $\xi_{\rm h}=17.5$ and
$\xi_{\rm w}=70.5$. For this model ${\cal R}=10^{-3}$ yr $^{-1}$ and
the rate of mass deposited by SNe is $\dot M_{\rm SN}=10^{-2}$ $\msun$
yr$^{-1}$. Thus, if the rate of mass transfer from the clouds to the
diluted gas is larger than $\sim 0.7$ $\msun$ yr$^{-1}$, the system
cannot seat on a steady solution and the HE is expected to be very
low.  For cloud mass transfer rates smaller than $\dot M_{\rm
SN}=10^{-2}$ $\msun$ yr$^{-1}$ but larger than $\sim 0.17$ $\msun$
yr$^{-1}$, both evolutions toward high values or toward low values of
the HE are possible.  It is noteworthy that, for a SN rate 100 times
larger, as that estimated for M82, we have $\dot M_{\rm SN}=1$ $\msun$
yr$^{-1}$ and $\xi_{\rm w}=14.3$. In order to obtain low values of the
HE, clouds should lose mass at a rate greater than 14 $\msun$
yr$^{-1}$.  As shown below, such a high rate can be sustained only for
short times compared to the SB age, $t_{b}$. The value of the HE is
thus expected to be close to unity in this case, as suggested by
Chevalier \& Clegg (1985).

As shown by this simple example, the behavior of the ISM in the SB
region is regulated by the value of $\xi$. Thus, in order to make a
more realistic model, we must take into account in more detail the
mass exchange mechanisms between clouds and diffuse ISM. They include
SNRs interactions, clouds-SNR interactions, photoevaporation, drag due
to Kelvin-Helmholtz instability and thermal evaporation.  We discuss
these mechanisms in the next section.
\section{Physical processes in the SB environment}

\subsection{Clouds formation}

Cold dense gas is observed in SB regions, as in M82 (Kerp, Herbstmeier \&
Mebold 1993;
Cecil et al. 2001; Pietsch et al. 2001 ).  Radio observations of SB
galaxies indicate the presence of clouds with masses between $10^2
M_{\odot}$ and $5 \times 10^3 M_{\odot}$, radii between 0.5 pc and 1
pc, and temperatures between 50 K and 200 K (e.g. Cesaroni et
al. 1991; Carral et al. 1994; Paglione, Tosaki \& Jackson 1995;
Garay \& Lizano 1999).
However, we could also expect the presence of even smaller clouds
which are not detectable due to observational limitation. In our own
Galaxy, molecular clouds are detected in a range between 1 $M_{\odot}$
and 10$^6 M_{\odot}$ with a mass spectral index distribution $\alpha$ = 1.5
(Scalo \& Lazarian 1996). These molecular clouds show in turn
sub-structures called clumps, that are distributed in a mass range
between $10^{-4}$ and $10^3$ M$_{\odot}$, with power law spectral
index for the distribution in a range $\alpha$ = 1.4-1.9 (Kramer et
al. 1998; Blitz 1993).

While most of the cold gas observed in SB regions is pristine, new
cold clouds may  form under the action of SNRs.
Cioffi and Shull (1991) have shown that many small clouds
(with sizes of fractions of parsec) form after the interaction among SNRs
generated by a high rate of SNe explosions.
Scalo \& Chappell (1999) highlight the
formation of cold filaments in HII regions, as a consequence of strong
stellar winds and SNe explosions occurring in a region containing a
large number of stellar sources.

\begin{figure}
\centering \epsfxsize=7cm \epsfbox{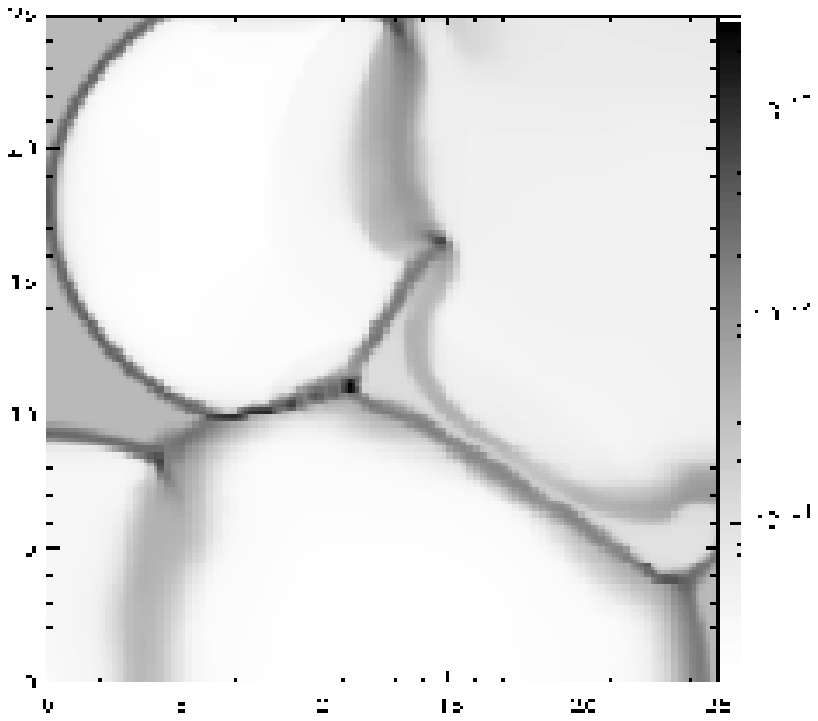} \epsfxsize=7cm
\epsfbox{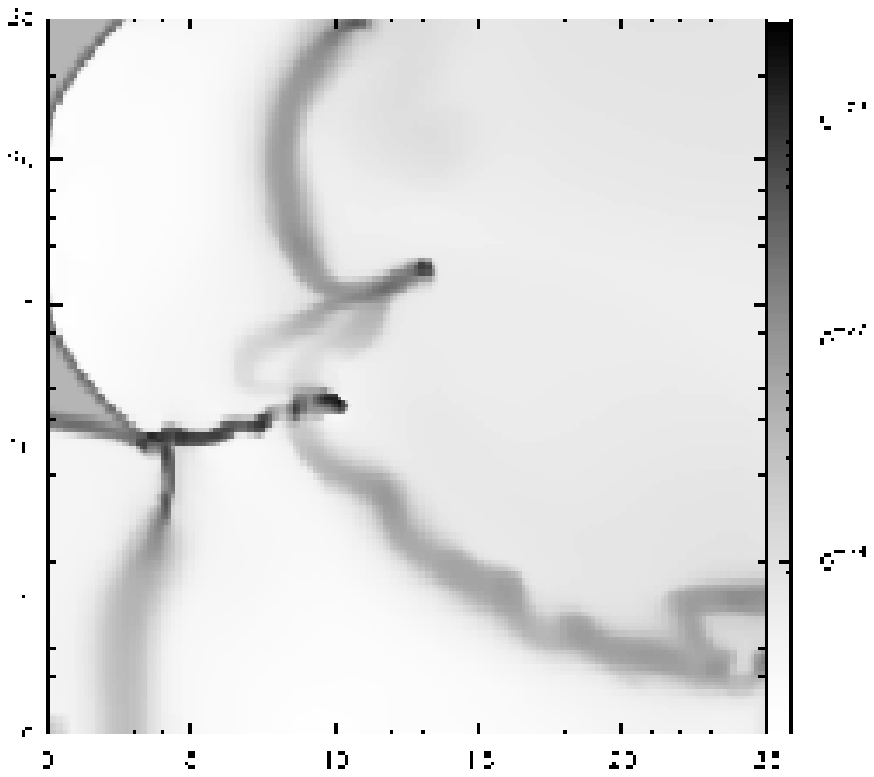} \caption{Grey-scale map of the  density
distribution showing the interaction of 5 SNRs in a box of 25 pc
$\times$ 3.125 pc $\times$ 25 pc, corresponding to 256 $\times$ 32
$\times$ 256 grid points at the highest grid level. The left
and bottom faces that define the beginning of the computational
box have reflective boundaries and the right and top faces have
continuous (or outflow) boundaries. The ambient medium has $n$=1
cm$^{-3}$ and T= $10^4$ K. The initial energies of the SN
explosions are assumed to be: $E_0$ = $6.3 \times 10^{49}$ erg,
$1.6 \times 10^{50}$ erg, $4.2 \times 10^{49}$ erg, $3.8 \times
10^{49}$erg, and $1.3 \times 10^{50}$ erg. The panels
correspond to two different snapshots of the evolution of the
system: $t$ = $2.5 \times 10^4$ yr (upper panel) and $4.5
\times 10^4$ yrs (lower panel) (see also Melioli, de Gouveia 
Dal Pino \& Raga 2004, for details).}
\end{figure}

We have recently run three-dimensional (3D) hydrodynamical simulations 
of SNRs interactions employing  a modified  version of the  Yguaz\'u-a 
adaptive grid code originally developed by Raga, Navarro-Gonzalez \&
Villagran-Muniz (2000; see also Raga et al. 2002,
Masciadri et al. 2002, Gonzalez et al. 2004). This code
integrates the hydrodynamic equations explicitly accounting for the radiative
cooling together with a set of continuity equations for several atomic/ionic
species employing the flux-vector splitting algorithm of Van Leer (1982).
The following species have been considered: H I, H II,
He I, He II, He III, C II, C III, C IV, O I, O II, and O III.
The reaction
rates and the non-equilibrium cooling function are given in Raga et al. (2002).
The calculations were performed on a four-level, binary adaptive grid with a
maximum resolution along the x, y, and z axes of $0.1 $ pc. The
computational domain extends over ($25) \times (3.125) \times 25$
pc, corresponding to $ \times 256 \times 32 \times 256$ grid points
at the highest resolution
grid level (see also, Melioli, de Gouveia Dal Pino \& Raga 2004 ).

The simulations have shown that the formation of very dense filaments (1000
times denser than the hot gas) and dense surfaces (100 times denser
than the hot gas) is possible in times
$<$10$^5$ yrs (Figure 3).  Rayleigh-Taylor (R-T) and  Kelvin-Helmholtz 
(K-H) instabilities are able to fragment these structures and produce new
clouds, in a process that can continue during all the SN activity.
\footnote{An alternative mechanism for clouds formation could be due to
ISM turbulence (e.g., Ehlerova et al. 1997, Palous, Wunsch \& Ehlerova 2001, 
Ortega, Volkov \& Monte-Lima 2001, Elmegreen 2003). Dynamical instabilities 
and stellar pressures can drive turbulence in the ambient medium that in turn 
may cause the formation of clouds and filaments. However, as most of these 
studies relying on turbulence have neglected the presence of SN shell 
interactions at a high rate, the resulting timescales for clouds formation 
and their resulting sizes are larger than those expected in the SB 
environment.}

We assume that the rate of clouds formation depends on the instantaneous
amount of total mass of gas in the SB region.
We can express the rate of formation of the new clouds as:

\begin{equation}
 {dN_{c,int}\over dt}={\delta M_{g,h} \over {m_c} t_{int}}
\end{equation}
\noindent
where $N_{c,int}$ is the total number of clouds formed by SNRs shell
interaction, $M_{g,h}$ is the total mass of gas that fills the SB
volume, $m_c$ is the cloud mass, $t_{int}$ is the SNR interaction time
(see also $\S$ 3.4 below), and $\delta$ is the fraction of the gas
that is deposited in a SNR shell. Models for the evolution of a SNR (see,
e.g., Chevalier 1974) indicate values for $\delta$ between 0.9 and
0.99.

\subsection{Photoevaporation}
The SB volume is filled by a huge amount of ionizing photons which are
produced by the massive stars. In order to estimate the number of ionizing
photons in our SB system, we have run the ``mappings'' code developed by
Leitherer et al.  (1999), which is available in the site {\it www.stsci.edu/
science/ starburst99/ mappings}. Using initial conditions which are
appropriate to our SB environment, we have found that
the SN phase begins at a time of 3.61 $\times 10^6$ yrs, that we have taken
as the zero-time in our model. At this time, the number rate of ionizing
photons is $S=2.25\times 10^{52} s^{-1}$, and after this time it decreases
as $t^{-4}$.

Initially, when a cloud becomes exposed to the ionizing radiation, it
implodes and injects part of the mass into the ambient medium (Bertoldi 1989).
A compressed neutral globule remains, which continues to evaporate by
photoionization (Bertoldi \& McKee 1990).

If we take into account the presence of a non-zero magnetic field in the cloud,
this can increase its effective pressure and make it more resistant against
photoevaporation.
In fact, in the case of the ISM of our own Galaxy,
the pressure balance of the clouds with the hot high pressure, diffuse
component
is often dominated by the magnetic pressure rather than the thermal pressure
of the cloud (e.g., Cox 1995, Bower et al. 1995).
In the case of a magnetically dominated cloud, it can be
shown that when the pressure exerted by the ionizing front balances the
pressure of the neutral globule cloud, then (Bertoldi \& McKee 1990):
\begin{equation}
{p_{\rm c}\over k}=2.75\times 10^8\left ({n_{c} \over
1000}\right )^{1/21} {1 \over {b^{2/7}}} \flx^{4/7}m'^{-4/21}\;
{\rm cm^{-3}\,K},
\end{equation}
where $m'$ is the cloud mass in $M_{\odot}$, $n_{c}$ is the
initial cloud gas density and the parameter $b$ gives the initial magnetic
field expressed as $B_0=bn_c^{1/2}$ $\mu$ G, and $\flx=S_{49}/ {R_{SB}}^2$
with
$S_{49}=S/10^{49}$ s$^{-1}$.
In our reference SB model (see below) we obtain $\flx=0.225$ pc$^{-2}$
s$^{-1}$.
For a magnetically dominated cloud  $b \geq 0.3$, and the photoevaporation
rate can be written as a function of the cloud mass as (Bertoldi \& McKee
1990):
\begin{equation}
\dot M_{UV}=-1.79\times 10^{-5} \psi \flx ^{2/7}m'^{4/7} \;\;\msun \;{\rm
yr^{-1}},
\end{equation}
\noindent
where $\psi$ is a dimensionless factor:
$$\psi \sim {{2.7 \ b^{6/7}} \over {n_c^{1/7}}}$$
where $n_c$ is given in cm$^{-3}$.
In this work, in the presence of photoevaporation, we have $n_c \simeq 10^4 -
6\times 10^5$ cm$^{-3}$, which implies $\psi=0.5-1$ (see $\S$) for b$\sim$1.

Finally, it can be shown that the radius of the cloud is:
\begin{equation}
r_c=2.6\times 10^{-2}\flx^{-1/7}m_c^{8/21}\;{\rm pc}.
\end{equation}
\noindent
\noindent

It is worth noting that the O and B stars are able to
photoionize the clouds only if these are closer than the Str\"{o}mgren
radius
\begin{equation}
R_s = 66.9 ({{S_{49}} \over n^2})^{1/3} \ \ \ \ {\rm pc}.
\end{equation}
\noindent
Assuming a number of O and B stars $\sim {\cal N} /3$, the filling factor of
the photoionizated SB volume, $ff_s$, is:

\begin{equation}
ff_s = 3 \times 10^5 {{\cal N} \over {3 R_{SB}^3}} ({{S_{49}} \over n^2}).
\end{equation}
\noindent
Thus, the cloud mass-loss rate due to photoionization must be scaled by the
factor $ff_s$, if $ff_s < 1$.
We will see below, however, that $ff_s$ increases to 1 in a quite short time
($\leq 2 \times10^6$ Myr) compared to the duration of the SN activity.

\subsection{Drag}

The cloud motion through the medium leads to mass-loss by the
clouds due to the Kelvin-Helmholtz (KH) instability (e.g. Lin \& Murray 2000).
Defining $\chi \equiv \rho_{\rm c}/\rho$ as the gas density ratio
between the cloud and the diffuse ISM, the growth timescale of the K-H
instability is given by (Klein, McKee \& Colella 1994)
$\tau_{\rm KH}\sim \rc \chi^{1/2}/\vc$.
Thus, the mass-loss rate $\mc/\tau_{\rm KH}$ can be written as
\begin{equation}
\dot M_{\rm d}\sim 1.3\times 10^{-6} n \chi^{1/2} r_c^2
v_{\rm c,6} \;\;\msun\;{\rm yr}^{-1},
\end{equation}
\noindent
where $v_{\rm c,6}$ is the cloud velocity in units of $10^6$ cm s$^{-1}$ and
$r_c$ is expressed in pc.
Taking into account equation (8), we may express this rate as a function of
the cloud mass:
\begin{equation}
\dot M_{\rm d}\sim 8.8\times 10^{-10}n \chi^{1/2} v_{\rm c,6}
\flx^{-2/7}m_c^{16/21} \;\;\msun\;{\rm yr}^{-1}.
\end{equation}
\noindent
where $\chi$ is obtained assuming that the outer shell of the clouds is at a
temperature $\sim 10^4$ K.

\subsection{SNR-clouds interactions}

Besides the classical drag, clouds may also be destructed by
interaction with SNRs. This is a rather complex process, and
its investigation deserves numerical simulations.
Works in literature dealing with this problem are rare.
Poludnenko, Frank and Blackman (2002) have recently performed 2-D
simulations of the
interaction of a strong, planar shock wave with a set of dense inhomogeneties,
embedded in a more rarefied ambient medium. Their results show that
after compression, re-expansion and destruction, a mixing phase occurs
that causes a high increase of the ambient density.

More recently, motivated by the investigation in the present work, we
have carried out fully 3-D hydrodynamical radiative cooling
simulations of the interaction of a SNR shell with dense clouds
embedded in a hot gas. As described before, the
 simulations were performed with an unmagnetized version of the
adaptative grid code YGUAZU (Raga, Navarro-Gonzalez \&
Villagran-Muniz 2000; Raga et al. 2002;
Masciadri et al. 2002) which solves the gas dynamical equations
together with a set of continuity equations for several atomic/ionic
species without including the effects of photoevaporation and thermal
evaporation of the clouds.  With the radiation rates and
non-equilibrium cooling function given by Raga et al. (2002) (see also
Melioli, de Gouveia Dal Pino \& Raga 2004 for details) our 
results, somewhat distinct from
those of Poludnenko, Frank and Blackman (2002), show that after a time
between $\sim 10 ^4$ and $10^5$ yr the clouds are largely fragmented. However,
little mixing occurs between the mass lost by the clouds and the ISM;
instead, we find that new filaments form and the shell mass grows if
the initial density of the clouds is of the order of that of the shell
(see Fig. 4). Only in few cases, when there are several very little
clouds (with $r_c \ll h_{sh}$ and $n_c \geq n_{sh}$, where $h_{sh}$
and $n_{sh}$ are the thickness and the number density of the shell,
respectively) the mixing occurs and the ISM density effectively grows
by a factor 10-100 times larger than that of the hot gas density.
Thus, as a rule, this type of interaction does not affect the diffuse
ISM global density but contributes to form more smaller clouds,
fragmented shells and, consequently, new generations of clouds.  In
this work, where we have assumed clouds with typical radius $r_c \leq
h_s$, the fragmentation time due to these interactions is:
\begin{equation}
t_{d,sh} \sim {h_{sh} \over {(v_{sh}-v_c)}} \sim 10^4 {\rm yr}
\end{equation}
\noindent
with $v_c$ and $v_{sh}$ being the clouds and the shell velocity, respectively,
which is in agreement with the simulations.
{\footnote {We notice that in the simulation of Fig. 4, we have taken clouds
with densities $n_c$=100 cm$^{-3}$; for larger densities (as those we have
considered in $\S$ 5) we must expect a similar effect, i.e., an efficient
fragmentation of the clouds but without significant increase of the ambient
gas density.}}

\begin{figure}
\centering
\epsfxsize=9cm \epsfbox{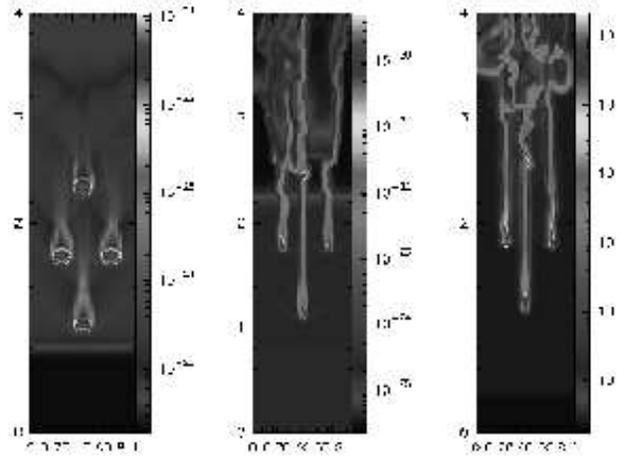} \caption{Color-scale maps
showing the time evolution of the midplane density distribution
(in log scale) of the interaction of two SNR shells with a system
of 4 clouds with $m_c = 0.01 M_{\odot}$, $r_c$=0.09 pc, and
$n_c$=100 cm$^{-3}$. The shells are injected at  times $t$ = 0 and
$t$ = 4.5 $\times 10^4$ yr. Each shell has a density $n_{sh}$=0.09
cm$^{-3}$ and velocity of 250 km ${\rm s^{-1}}$, and the ambient
medium has $n$=0.01 cm$^{-3}$ and T= $10^6$ K. The computational
box has dimensions 4 pc $\times$ 1 pc $\times$ 1 pc, corresponding
to 512 $\times$ 128 $\times$ 128 grid points at the highest grid
level. Outflow boundaries have been assumed.  The times
depicted are: $t$ = 1.3 $\times 10^4$ yr (left), $t$ = 5.7
$\times 10^4$ yr (center), and $t$ = 8.6 $\times 10^4$ yr (right).
The density is shown in units of g cm$^{-3}$ (see Melioli, de Gouveia 
Dal Pino \& Raga 2004 for details).}
\end{figure}

Kinetic energy is also transmitted by SNRs to the clouds, which is lost
by cloud-cloud collisions. This balance is quite difficult to
simulate numerically. In our model we simply assume that the cloud
velocity stays constant at the fiducial value 10 km s$^{-1}$.

\subsection{Thermal evaporation}
A cold and dense cloud surrounded by a tenuous and hot gas of number
density $n$ and temperature $T$ may undergo evaporation at a mass rate
given by (Cowie, McKee and Ostriker 1981):
\begin{equation}
\mde\,({\rm \msun\; yr^{-1}})=\cases{-4.34\times 10^{-7}\phi T^{5/2}_6 r_c &
if $\sigma < 1$\cr
                            -5.92\times10^{-7}\phi T^{5/2}_6 r_c
\sigma^{-5/8} & if $\sigma >1$\cr 1.3\times 10^{-8}T^{5/2}_6 r_c
\sigma^{-1} & if $\sigma<\sigma_{\rm cr}$ \cr}
\end{equation}
\noindent
where $r_c$ is in pc, $T_6 = T/10^6 $ K, the factor $\phi < 1$ allows
for a reduction in the heat conduction due to the presence of magnetic
fields,
turbulence, and other effects,
{\footnote {In the present work, due to the lack of knowledge of the
 magnetic field configuration in the clouds in a  SB environment  and thus
of its truly effect upon the heat transport, we have for
simplicity adopted $\phi =1$. Although this will tend to maximize the thermal
evaporation rate of the
clouds, we find that for the typical conditions of the clouds in the SBs the
evaporation will, in general, be  dominated by the photoevaporation ablation
process (see $\S$ 4 and Fig. 6 below).}}
and
\begin{equation}
\sigma=\left ({T \over 1.54\times 10^7\,{\rm K}}\right )^2 {1\over n\rcpc \phi}
\end{equation}
is the saturation parameter.
For $\sigma<\sigma_{\rm cr}$ clouds condense rather than evaporate (McKee \&
Cowie 1977; McKee \& Begelman 1990).
These latter authors find $\sigma_{\rm cr}=0.028$, but Borkowski, Balbus \&
Fristrom (1990) claim that the value should be five times larger:
$\sigma_{\rm cr}=0.15$.

Also in this case the mass-loss rate can be obtained as a function of
the cloud mass $\mc$ whith the help of equation (8). In the discussion in the
next section, we focus on the classical case ($\sigma <1$), which
represents an upper limit to the mass evaporation rate. In this case:
\begin{equation}
\dot M_{ev} = {-6.94 \times 10^{-9}\phi T^{5/2}_6
\flx^{-1/7}m_c^{8/21}}\;\;{\rm \msun\; yr^{-1}}.
\end{equation}

\section{Time scales}

The energy released by the SNRs and the evolution of the ISM will
depend on the shells interaction time, $t_{int}$.

A SNR will
form only after the SN shock front enter the Sedov phase.
When the SN explodes, the star ejects a mass $M_{ej} \simeq 10$ M$_{\odot}$
into the ISM with a terminal velocity $\sim 10^4$ km s$^{-1}$ and
the ejecta will expand at nearly constant velocity until they encounter a
comparable mass of ambient medium. This occurs at a time $t_{sh}$ which
determines the onset of the SNR formation  (e.g., McCray 1985):

\begin{equation}
t_{sh} =  {200 \over {n^{1/3}}} ({M_{ej} \over M_{\odot}})^{1/3} \ \ \ \
{\rm yr}
\end{equation}
where n is ambient number density.
The present model will be valid only for $t_{sh} < t_{int}$.

To compute the
 shells interaction time, $t_{int}$, we will borrow the $porosity$ concept
first introduced to investigate the possibility that the collective effects
of the SNRs in our own Galaxy might be capable of generating and maintaining
a rather widespread component of much hotter, lower density gas in the ISM
(Cox \& Smith 1974).
Here, we will assume that $t_{int}$ corresponds
to a porosity value (Cox \& Smith 1974, McKee, 1992):
\begin{equation}
Q = {{\rat V_{SN} t_{int}} \over {V_{SB}}} \sim 1
\end{equation}
\noindent
where $V_{SN}$ is the volume of a SNR at the time $t_{int}$ and $V_{SB}$ is
the volume of the SB ambient.
$Q$ gives, approximately, a measure of the hot gas filling factor in
space-time.
 When it is close to unity, all the ambient medium must be filled by SNRs.
The shells collide with each other and a unique superbubble of hot gas
will fill the SB region and may drive a galactic superwind.
{\footnote {Although there has been some debate in the literature
regarding the uncertainty in the determination of the value of the porosity
induced by the remnant population in the ISM of our  Galaxy and its real
importance in providing and maintaining a hot gas component
(Cox 1995, McKee 1995; see also Slavin \& Cox 1992, 1993, McKee \& Zweibel
1995), in the case of SB galaxies, the SN
rate is more than 10 times higher than in normal galaxies and therefore, SNRs
interactions are expected to  be efficient enough to eventually provide a
substantial value for the porosity.}}

Shells interactions can occur before or after the SNRs become radiative.
Depending on this, we have two possible expressions for $t_{int}$.
If the shells interact when the SNRs are still in the Sedov phase:
\begin{eqnarray}
t_{int-sedov} = 9 \times 10^4 \ {({{100 \ {\rm {pc}}} \over R_{SB}})}^{15/11}
\nonumber
\end{eqnarray}
\begin{equation}
\ \ \ \ \ \ \ \ \ \ \ \ \ \ \ \ \times ({{3.3 \times 10^{-4} \
{\rm {yr^{-1}}}} \over {r_{sn}}})^{5/11}
{({n \over E_{51}})}^{3/11}
\ \ {\rm {yr}}
\end{equation}
\noindent
and if the shells interact when the SNRs are already in the radiative phase,
then:
\begin{eqnarray}
t_{int-rad} = 7.5 \times 10^4 \ {({{100 \ {\rm {pc}}} \over R_{SB}})}^{21/13}
\nonumber
\end{eqnarray}
\begin{equation}
\ \ \ \ \ \ \ \ \ \ \ \ \ \ \times ({{3.3 \times 10^{-4} \
{\rm {yr^{-1}}}} \over {r_{sn}}})^{7/13}
{n}^{0.42} {{\beta ^{0.06}} \over {{E_{51}}^{0.37}}} \ \ {\rm {yr}}
\end{equation}
\noindent
where $\beta$ has been defined in $\S$ 2.2 .
Figure 5 compares these interaction times (as functions of the ambient density)
with the time scale for the shell formation, $t_{sh}$, and with the radiative
cooling time of the SNR shell (McCray 1985):
\begin{equation}
t_c = 1.49 \times 10^4 {\beta}^{-5/14} n^{-4/7}
\end{equation}
\noindent
We notice that for ambient densities $\leq 0.1$ cm$^{-3}$, we must take
$t_{int} = t_{int-sedov}$, since $t_{int-sedov} \leq t_c$ in this range.
For $t \geq t_c$, the SNR enters the radiative phase and $t_{int-rad}$ should
be employed.
Since for most of the calculations of interest in this study (see below), we
take either $n=0.01$ cm$^{-3}$ or 5 cm$^{-3}$, we will use
$t_{int}=t_{int-sedov}$ or $t_{int}=t_{int-rad}$, respectively, to evaluate
the time scale for superbubble formation.
\begin{figure}
\centering
\epsfxsize=8cm
\epsfbox{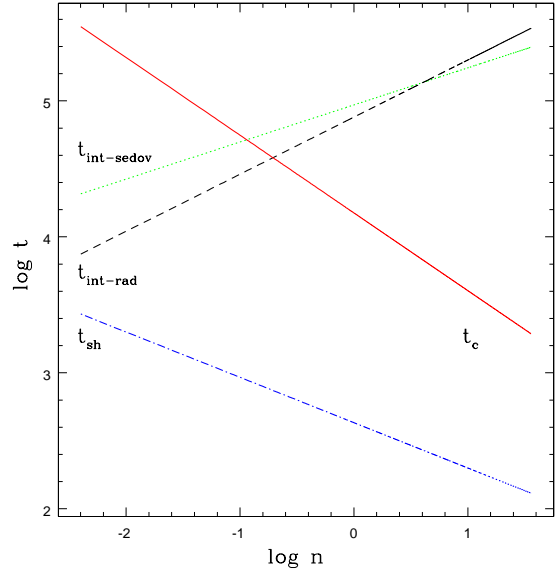}
\caption{Time scales versus gas ambient density of a SNR in a spherical SB
region, with $R_{SB}$=100 pc.
The interaction time in the Sedov phase is represented by the dotted
line, the interaction time in the radiative phase by the dashed line, the
time scale for the radiative cooling by the solid line, and the time scale
for the formation of the shell by the dot-dashed line.}
\end{figure}

Using equations 7, 12 and 16, and defining the time scale for the ambient
density increase
as the time a given cloud mass-loss process (photoevaporation, thermal
evaporation or drag) takes to increase the ambient number density by 1
cm$^{-3}$, then we have:
\begin{equation}
t_{UV}= 837 ({R_{SB} \over {100 \ {\rm pc}}})^{3} ({{8 \times 10^6 \
M_{\odot}} \over M_g}) {{{m_c}^{3/7}} \over {{\psi} \flx^{2/7}}} \ \ {\rm {yr}}
\end{equation}
\noindent
\begin{equation}
t_{ev}= 1.9 \times 10^4 ({R_{SB} \over {100 \ {\rm pc}}})^{3}
({{8 \times 10^6 \ M_{\odot}} \over M_g}) {{{m_c}^{2/3} n_c^{1/3}} \over
{{T_6}^{5/2}}} \ \ {\rm {yr}}
\end{equation}
\noindent
\begin{equation}
t_{d}= 2.9 \times 10^3 ({R_{SB} \over {100 \ {\rm pc}}})^{3} ({{8 \times
10^6 \ M_{\odot}} \over M_g}) {{n_c^{2/3}} \over {v_6}} {{m_c^{1/3}} \over
{n \ {\chi^{1/2}}}} \ \ {\rm {yr}}
\end{equation}
\noindent
for the density increase timescale through photoevaporation, thermal
evaporation and classical drag, respectively.
In the equations above, $n$ and $n_c$ are given in cm$^{-3}$.
Depending on how all these time scales relate to each other, one finds either
a superbubble scenario or a gas mixing scenario.
When $t_{int}$ is longer than one of the three time scales above,
the mixing of the ablated matter from the clouds occurs
before the formation of a high pressure bubble.
In this case, the ablated gas will enhance the density of the hot diffuse gas
of the ISM and increase its radiative cooling.
In contrast, if $t_{int}$ is shorter than the destruction time scales
above, a low density superbubble will form very fast and the SN heating
efficiency (HE) will rapidly increase to unity.

Figures 6 compares these time scales for typical parameters of a SB
region, where the time scales are depicted as a function of the mass of the
clouds.
We see that for the two adopted pair of values of the density and temperature
of the ambient medium, $n=0.01$ cm$^{-3}$ and $T=2 \times 10^6$ K, and $n=5$
cm$^{-3}$ and temperature $T= 5 \times 10^4$ K, the presence of clouds with
mass $m_c \leq$ 100 $M_{\odot}$ will make  $t_{int}$ longer than at least
one of the time scales for clouds ablation.
In this case, the gas mixing scenario will be dominant and the formation of a
superbubble is postponed.

\begin{figure}
\centering
\epsfxsize=8cm
\epsfbox{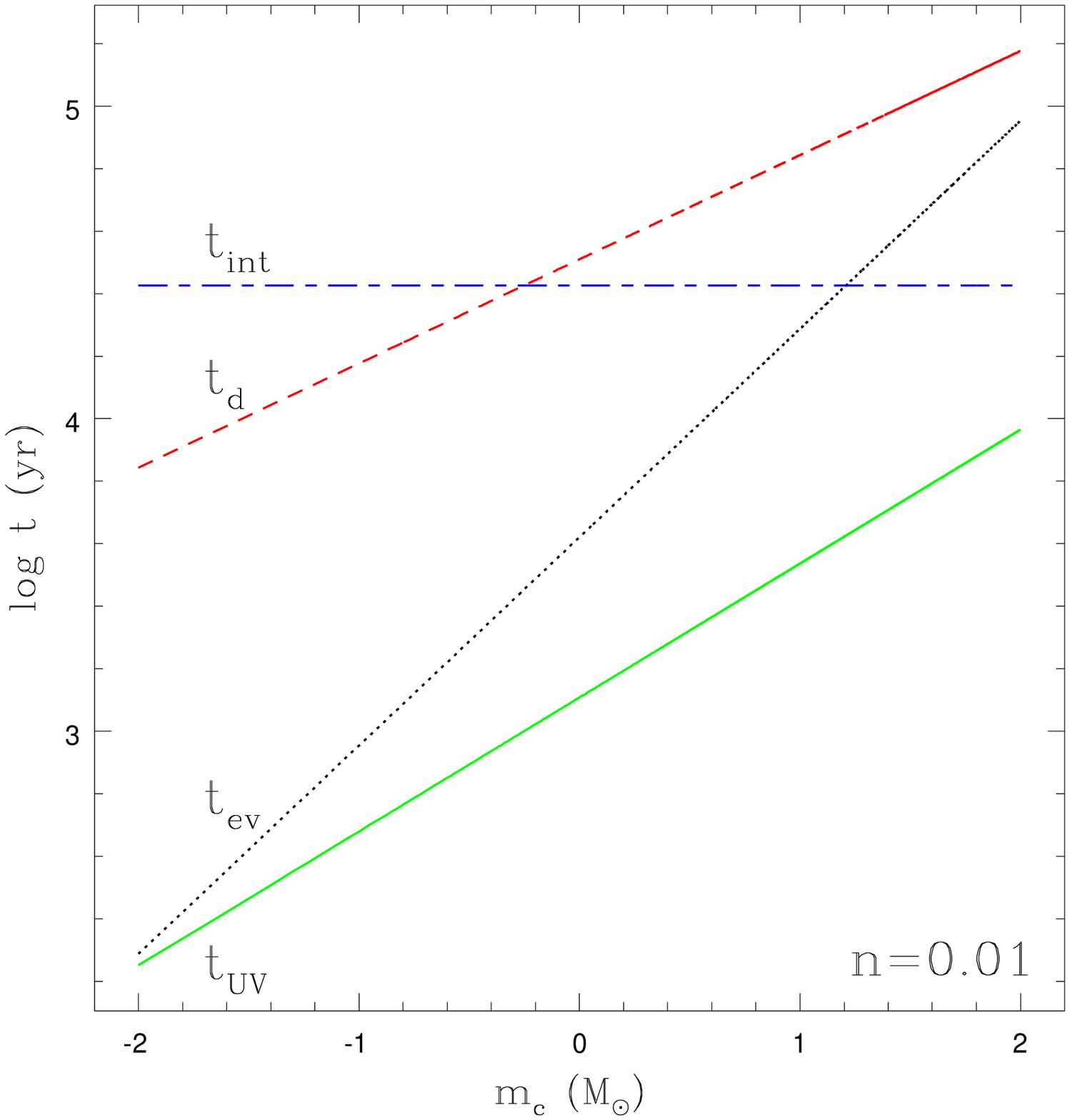}
\epsfxsize=8cm
\epsfbox{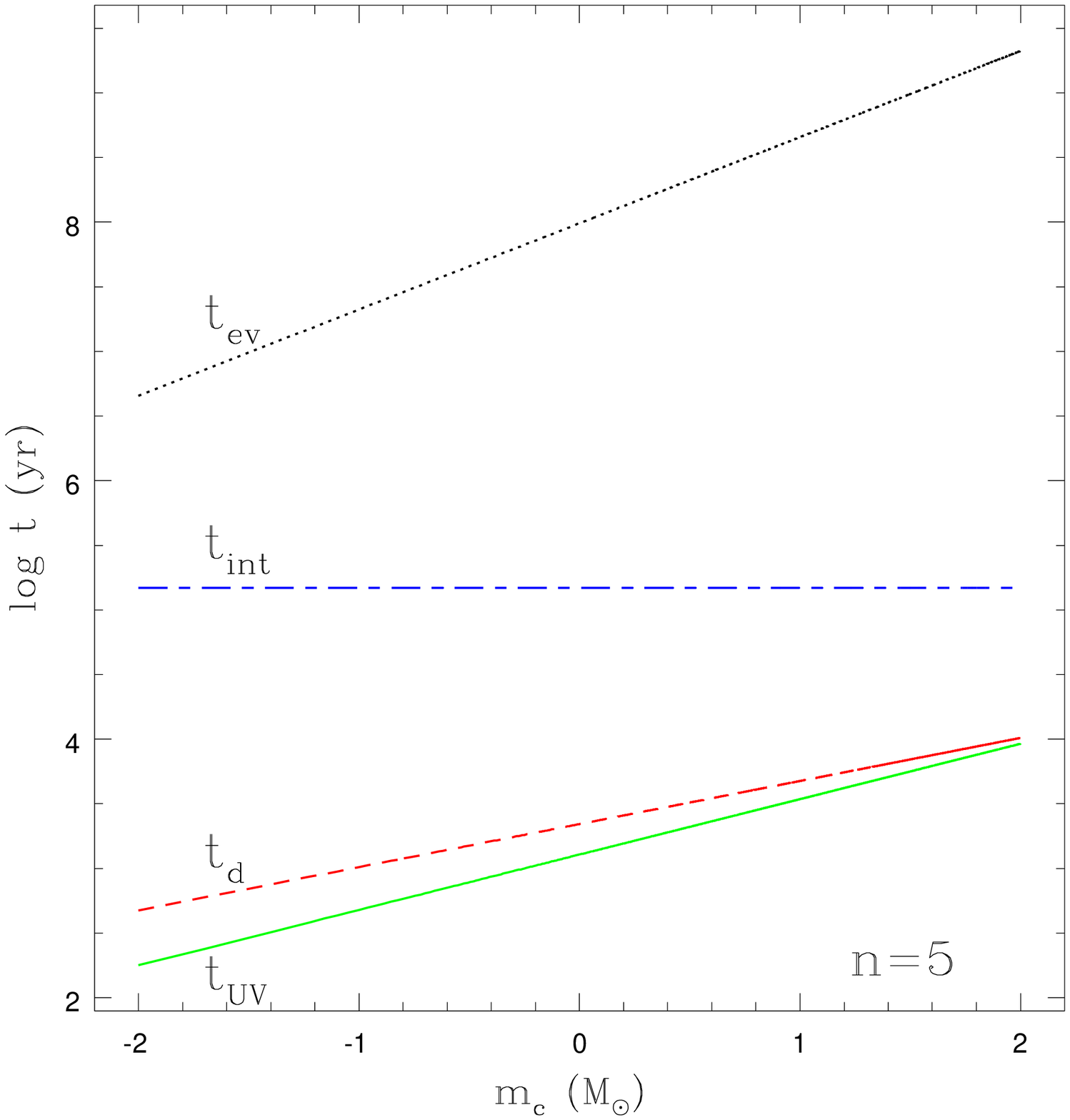}
\caption{Steady time scales versus mass of
the clouds in a spherical SB region, with $R_{SB}$=100 pc, $\flx$=0.225, $M_g
=8 \times 10^6 M_{\odot}$, $\psi$ = 1.
The interaction time is represented by the dot-dashed
lines, the ambient density increase timescale due to photoevaporation by
solid lines, due to thermal evaporation
by dotted lines, and due to drag by dashed lines.
Upper panel: $n$=0.01 cm$^{-3}$, T=$2 \times 10^6$ K; bottom panel:
$n$=5 cm$^{-3}$, T=$5 \times 10^4$ K.}
\end{figure}

\section{An evolution model for HE}
\subsection{Equations}
Once considered all the relevant physical processes, we can now write
the equations governing the evolution of our model described in
section 2.1. In particular, we describe the time evolution of the
ambient gas density, $n$, and pressure, $p$, and of the clouds number
$N_c$ with a mass $m_c$. The set of equations is given by:

\begin{equation}
{{d M_{loss}} \over dt} = \dot M_{UV} +\dot M_{ev} +\dot M_d \ \
({\rm M_{\odot} \ yr^{-1}})
\end{equation}
\begin{equation}
{dN_c \over dt} = C_{1}{n \over t_{int}} -
           N_c{\dot M_{loss} \over m_c} - C_2 N_c v_c -
           C_{3}\dot N_c {m_c \over t_{coll}} \ \
({\rm yr^{-1}})
\end{equation}
\begin{eqnarray}
{dn \over dt} = C_4 N_c \dot M_{loss} - C_5{n \over t_{int}} -
C_2 n {\cal M} c_{\rm s}
\nonumber
\end{eqnarray}
\begin{equation}
\ \ \ \ \ \ \ \ \ + C_6 \ ({M_b \over 10^6}) \ \
({\rm cm^{-3} \ s^{-1}})
\end{equation}
\begin{eqnarray}
{dp \over dt} = C_7 {\cal R} E_{SN} + C_8 N_c \dot M_{loss} v_c^2 - C_9 n^2
\Lambda (T)
\nonumber
\end{eqnarray}
\begin{equation}
- C_{10}p {\cal M} c_{\rm s} \ \
({\rm erg \ s^{-1}})
\end{equation}
Eq. (25) computes the total mass-loss rate per cloud using the
equations of the clouds mass-loss rate (7), (12), and (16); Eq. (26)
gives the evolution of the total number of clouds; and Eqs. (27) and
(28) give the mass and energy conservation, respectively, of the
diffuse gas.  In these equations, $C_{\rm {n}}$ (with n = 1 to 10) are
constants which are given in Appendix B, $c_{\rm s}$ is the sound
speed of the ambient gas, $\cal M$ its Mach number, that is taken to
be equal to 1 in this study, and $v_c$ is the velocity of the clouds
which is assumed to be $10^6$ cm s$^{-1}$, as discussed in $\S$ 3.4.
In Eq. (26), the first two terms give the net difference between the
processes of cloud formation and cloud ablation, the third term gives
the rate of clouds escaping through the boundaries, and in the last
term we have introduced the typical time scale for cloud-cloud
collisions, $t_{coll}=1/(N_c 4 \pi r_c^2 v_c)$, in order to account
for the possibility of collisions between clouds in the SB volume.  We
note that we have assumed a constant mass for the clouds, $m_c$.  This
is equivalent to adopting a constant mass distribution function (or a
constant average mass), and is a valid assumption, as long as the
processes for cloud formation and destruction do not affect much the
average mass of the clouds but only their total number.

In Eq. (27), the first term gives
the ambient gas density increase due to the total mass-loss rate by the
clouds; the second term gives the density decrease due to the interaction
with SNR shells that sweep the ambient material; the third term represents the
gas escaping through the boundaries, and ${\rm C_6}$ gives the injection of
matter from stellar winds and SNe explosions (Leitherer \& Heckman 1995).

Eq. (28) computes the energy of the diffuse gas. The first term is the
rate of energy injection from the SNe, the second gives the kinetic
energy term of the gas lost by the clouds, the third gives the energy decrease
due to the radiative cooling and the last term gives the energy density
decrease due to the escape of the gas through the boundaries.

Using a fourth-order Runge-Kutta integrator, we determine the evolution of the
SN efficiency HE defined as (see section 2.1):
\begin{equation}
{\rm HE}= 4 \pi R_{SB}^2 {{({1 \over 2}\rho v^2+{5 \over 2}p)}\over
{{\cal R} \rm E_{SN}}} {\cal M} c_s.
\end{equation}
\noindent

\subsection{Results}
In sections 2 and 3, we have studied the physical processes that drive the
mass-loss rate of the clouds and these have been compared
with the process for superbubble formation.
We have found that, while the time scale for the increase of the ambient gas
density, $t_{mx} = min(t_{UV}, t_{ev}, t_d)$, is shorter than the time scale
for the superbubble expansion, $t_{int}$, the HE value must be small.
When, on the other hand, $t_{int} \leq t_{mx}$, then HE must become close to
unity.
If this is correct, one should expect an increase in the HE value with the
decrease of the ambient gas radiative cooling, which is $\propto n^2$.

\begin{table*}
\caption[1]{Values for the parameters of the models assumed
in this work}
{\small
\centerline{
\begin{tabular}{|c|c|c|c|c|c|c|c|}
\hline\hline
\noalign{\smallskip}
\hbox{Model} & \hbox{$n$ (cm$^{-3}$)} & \hbox{T (K)} & \hbox{Cloud mass
(M$_{\odot}$)}
 & \hbox{Cloud mass-loss} & \hbox{$M_g$ (M$_{\odot}$)} & \hbox{$M_b$
(M$_{\odot}$)} \\
\noalign{\smallskip}
\hline
\noalign{\smallskip}
\hbox{1} & \hbox{0.01} & \hbox{$2 \times 10^6$} &
\hbox{$0.015$} &
\hbox{$\dot M_{UV}$, $\dot M_{ev}$, $\dot M_{d}$} & \hbox{$8 \times 10^6$} &
\hbox{$10^6$} \\
\noalign{\smallskip}
\hline
\hbox{2} & \hbox{5} & \hbox{$5 \times 10^4$} &
\hbox{0.015} &
\hbox{$\dot M_{UV}$, $\dot M_{ev}$, $\dot M_{d}$} & \hbox{$8 \times 10^6$} &
\hbox{$10^6$} \\
\noalign{\smallskip}
\hline
\hbox{3} & \hbox{0.01} & \hbox{$2 \times 10^6$} &
\hbox{$0.015$} &
\hbox{$\dot M_{UV}$, $\dot M_{ev}$, $\dot M_{d}$} & \hbox{$8 \times 10^7$} &
\hbox{$10^7$} \\
\hline
\hbox{4} & \hbox{0.01} & \hbox{$2 \times 10^6$} &
\hbox{0.015} &
\hbox{$\dot M_{ev}$, $\dot M_{d}$} & \hbox{$8 \times 10^6$} &
\hbox{$10^6$} \\
\noalign{\smallskip}
\hline
\hbox{5} & \hbox{0.01} & \hbox{$2 \times 10^6$} &
\hbox{mass dist. func. $<m>$ = 0.3} &
\hbox{$\dot M_{UV}$, $\dot M_{ev}$, $\dot M_{d}$} & \hbox{$8 \times 10^6$} &
\hbox{$10^6$} \\
\noalign{\smallskip}
\hline
\hbox{6} & \hbox{0.01} & \hbox{$2 \times 10^6$} &
\hbox{0.015, no clouds formation} &
\hbox{$\dot M_{UV}$, $\dot M_{ev}$, $\dot M_{d}$} & \hbox{$8 \times 10^6$} &
\hbox{$10^6$} \\
\noalign{\smallskip}
\hline
\end{tabular}}
}
\end{table*}

\begin{figure}
\centering
\epsfxsize=8cm
\epsfbox{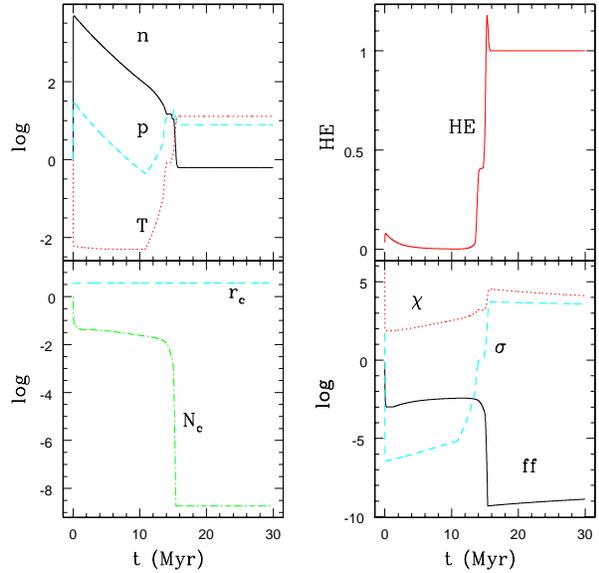}
\caption{Model 1: Ambient density ($n$), temperature ($T$) and pressure ($p$)
evolution in the top-left panel for a spherical SB with initial $M_b$ = $10^6$
M$_{\odot}$,
$R_{SB}$=100 pc, $n_0=0.01$ cm$^{-3}$, $T_0=2 \times 10^6$ K, $p_0 = 6.35
\times 10^{-12}$ dy cm$^{-2}$ and total gas mass $M_g=8 \times 10^6 M_{\odot}$.
In the bottom-left panel: number of clouds ($N_c$) and radius of clouds ($r_c$)
evolution;
in the top-right panel: the SN heating efficiency (HE); and in the
bottom-right panel: the clouds filling factor, ($ff=N_c {\cal V}_c/V_{SB}$,
where ${\cal V}_c$ is the volume of a single cloud), the saturation
parameter ($\sigma$, eq. 15) and the cloud to ambient gas density ratio
($\chi=\rho_c/ \rho$). The ambient density, temperature and  pressure, and the
number and radius of clouds are normalized to
 their initial values (in log scale).
The initial radius, temperature and density assumed for the clouds are $r_c=
6.5 \times 10^{-3}$ pc, $T_c=100$ K, and $n_c=6\times 10^5$ cm$^{-3}$,
respectively.}
\end{figure}

In Figure 7, we see the results for a reference model that we have called
Model 1. We have here considered neutral clouds ($T_c$=100 K) which are
initially in total pressure equilibrium with both the ambient gas and the
ionizing
photons.
Table 2 summarizes the input parameters for this model.
The SB has initial conditions which are taken from the limits given in
Table 1.
We notice that the SN heating efficiency increases to 1 after 16 Myrs.
Before this time, the total energy stored in the gas is only a few percent of
the total energy released by the SNe. Most of it is radiated away, as a result
of the heating of the dense gas.
The mass-loss rate from the clouds due to the processes of photoevaporation,
drag and thermal evaporation for Model 1 is shown in Figure 8. The
domination of the photoevaporation during the first 16 Myrs is clear, and
highlights the fact that after 16 Myr no clouds are present in the SB
(see bottom-left panel of Figure 7). Thus, after this time, the clouds
mass-loss becomes unimportant for the physical evolution of the system.
We also note that in this case, instead of thermal evaporation the clouds
actually suffer a thermal condensation, through $\dot M_{ev}$ (as $\sigma <
\sigma_{c}=0.03$, in the bottom-right panel of Figure 7), but this does not
influence the global results of the HE evolution.
\begin{figure}
\centering
\epsfxsize=8cm
\epsfbox{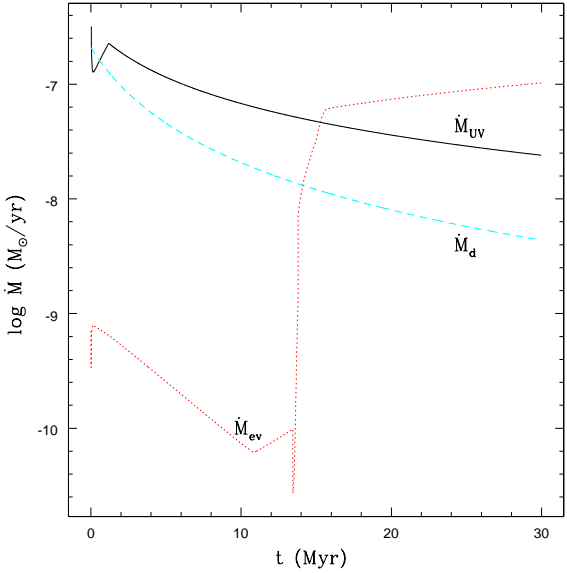}
\caption{Model 1. Mass-loss rate of the clouds with mass $m_c$ = 0.015
M$_{\odot}$ due to photoionization,
$\dot M_{UV}$ (solid line), thermal evaporation, $\dot M_{ev}$ (dotted line),
and drag, $\dot M_d$ (dashed line).}
\end{figure}

The initial large increase of the density (by four orders of magnitude) in
Figure 7, may seem unrealistic, in principle, but it is actually due to the
choice of the initial conditions.
Since we have run our model with an ISM with initial conditions which are very
similar to those of a superbubble ($n=0.01$ cm$^{-3}$, $T= 2 \times 10^6$ K)
and with $t_{mx} \leq t_{int}$,
Figure 7 shows that the ISM density increases almost instantaneously to
a typical value for SBs ($n \simeq 10-50 $ cm$^{-3}$).
Therefore, the assumed initial superbubble disappears before being able
to
blow out the gas.
Indeed, if we run a model with initial conditions that are typical of a SB,
i.e., with ambient $T=5 \times 10^4$ K and $n = 5 \ {\rm {cm^{-3}}}$
(see Figure 9, Model 2 of Table 2),
the density increases only by an order of magnitude, but the corresponding HE
value remains small during the same period of time.
\begin{figure}
\centering
\epsfxsize=8cm
\epsfbox{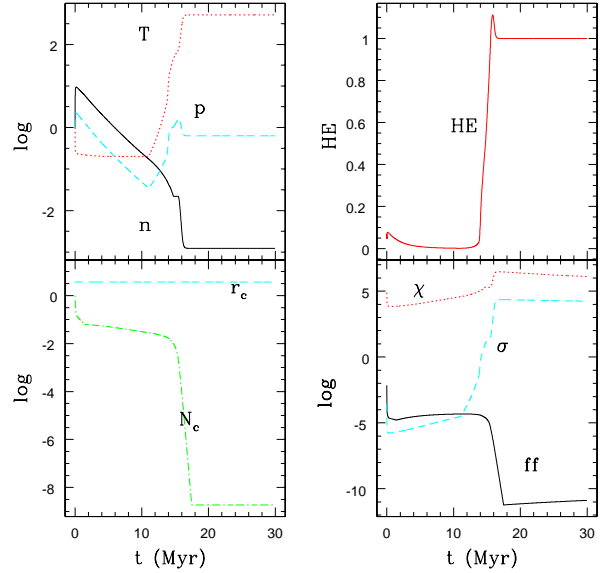}
\caption{Model 2. The same as in Model 1 of Figure 7, except that the initial
ambient gas density and temperature are $n$=5 cm$^{-3}$, $T=5 \times 10^4$ K,
respectively ($p_0 = 3.45 \times 10^{-11}$).}
\end{figure}

The comparison of the results of Figure 7 and 9 above indicates that the
HE
evolution is very similar in both models,
or, in other words, it is very insensitive to the initial conditions of the
hot ISM.

In both, as long as the ambient gas density remains large enough for the
radiative cooling to be efficient, the HE value remains very small.
After $\sim$ 16 Myr, when the density of the gas available in the SB becomes
too small ($\leq 10^{-1} {\rm {cm^{-3}}}$) due to continuous escape through
the boundaries of the system and exhaustion of the gas in the clouds,
then the radiative cooling ceases and a sudden increase in the
temperature leads to a superbubble formation in a time scale $t_{int} \simeq
10^5$ yrs. In consequence, HE $\rightarrow$ 1.

In Figure 10, we have run a model (Model 3 of Table 2) with the same
parameters of the Model 1, but with a total stellar mass 10 times bigger.
In our model, this implies an increase in the SN rate explosion ($\cal R$) and
the total gas mass in the clouds ($M_g$) of a factor 10.
Under these conditions, we see that HE never reaches high values during the
SB lifetime due to the larger gas mass in the clouds.
After an initial increase caused by the first explosion of the SNe at a higher
rate, the HE value decays to less than 0.5 per cent.

In Figure 11, Model 4 presents results for the HE evolution considering clouds
which are fully ionized ($T_c=10^4$ K). In this case, the photoevaporation was
not taken into account, since the clouds are transparent to the ionizing
photons. As a consequence, in this case they are in pressure equilibrium only
with the ambient medium.
The comparison with Model 1 of Figure 7 shows that the HE value increases
slower in Figure 11 (where HE $\rightarrow$ 1 at 20 Myrs),
because in the case of Figure 7 the clouds are more efficiently
destroyed, and survive for a smaller time in the SB.

At this point, it is interesting to highlight that the clouds are the real
valves that regulate the feeding of the diffuse ambient medium of the SB, in
the sense that they are able to simultaneously retain part of the gas to
themselves and to lose part of it to the ISM through cloud destruction,
thus maintaining the ISM density at high values, as long as
$t_{mx} \leq t_{int}$.

Using a cloud number distribution (which is given in Appendix A) $N(m_c)
\propto m_c^{- \alpha}$, with $\alpha$ = 1.5 and an
average cloud mass $<m> \simeq$ 0.3 $M_{\odot}$ (Model 5) instead of a
single initial cloud mass, the previous
results of models 1, 2 and 4 (for which a single cloud mass $m_c = 0.015$
M$_{\odot}$ was adopted) are not very much changed, as we see in Figure 12.
In this case, the density grows by 3 orders of magnitude at the beginning,
and the number of clouds evolve more smoothly than in
the other models due to the presence of a cloud mass distribution
instead of a single mass.
Nonetheless, since the density evolution is driven mainly by
the total amount of gas in the clouds, which is the same
as in the other models, the HE value goes to 1 after $\sim$ 16-19 Myr, i.e.,
just like in the previous models.
This result also shows that HE is quite insensitive to the adopted cloud mass,
$m_c$, since the average mass of Model 5 is almost 10 times larger than
the adopted cloud mass in the previous models ($m_c = 0.015$ M$_{\odot}$).
Consistently, tests made with a constant cloud mass $m_c$ equal to the average
mass of the distribution function above ($<m_c> \sim 0.3$ M$_{\odot}$), have
produced similar results to those of Model 1 of Figure 7.

All the models investigated above (Figs. 7 to 12), have assumed
 a continuous formation
of new clouds, during the SB lifetime, which are generated
by swept ambient gas accumulated on the SNR shells.
If we now consider an extreme situation in which no deposited gas
on the shells is allowed to form clouds (see model 6 of Table 2), then in this
case Figure 13
shows that HE quickly goes to one in about 2 Myr.
This situation is equivalent
to assume a fully ionized gas both in the shells and the ambient medium with
negligible radiative cooling, so that the time scale for clouds formation from
fragmentation of cooled material deposited on the shells is much longer
than the time scale for SNR interactions. In Figure 13 (model 6),
only the first initial population of cold clouds is present in the SB
environment and it is
soon ablated (mainly by photoevaporation) adding cold, dense gas to the
ambient
medium. However, this is not sufficient to prevent the fast overlap of the
SNRs
to form
a hot superbubble and
the temperature increase of the diffuse ambient gas to about  $10^7$ K. Figure
13 actually gives a $lower$ $limit$ for the time it would take for HE to
increase to one.
In a more realistic
scenario, at least part of the deposited material on the shells should form
new
generations of clouds, and HE would then increase to unity only after several
Myrs, like in Figs. 7 to 12.
We note further that, if we make all the gas, including that of the clouds,
initially fully ionized (like in Figure 11), so that no photoevaporation
occurs
afterwards and do not allow for formation of new clouds, like in Figure 13,
then
in this case, we obtain that the HE value goes to one only after $\sim$
8 Myr.

\begin{figure}
\centering
\epsfxsize=8cm
\epsfbox{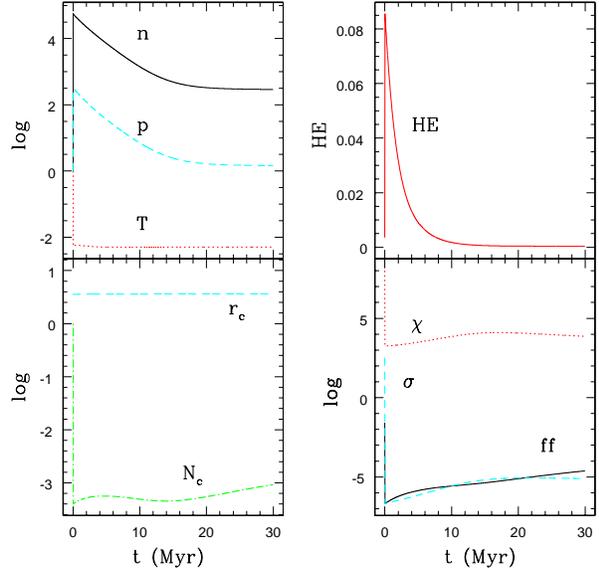}
\caption{Model 3. The same as in Model 1 of Figure 7, except that $M_b$ =
$10^7$ M$_{\odot}$ and $M_g = 8 \times 10^7$ M$_{\odot}$.}
\end{figure}

\begin{figure}
\centering
\epsfxsize=8cm
\epsfbox{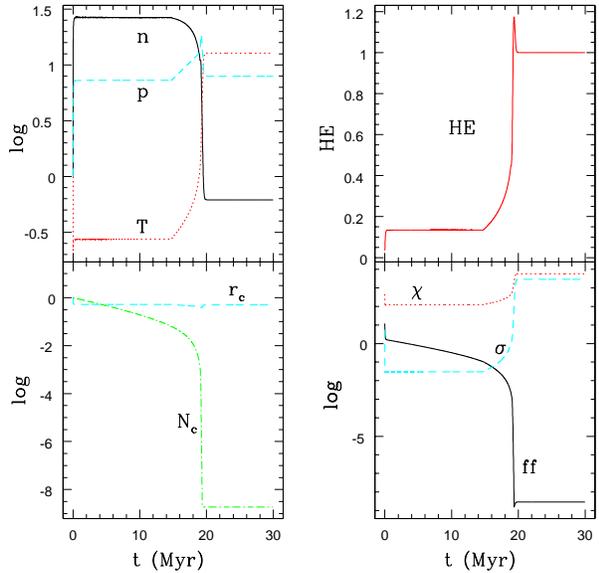}
\caption{Model 4. The same as in Model 1 of Figure 7, except that the clouds
are fully ionized ($T_c=10^4$ K, $r_c$=1.7 pc, and $n_c$=2 cm$^{-3}$), and
photoevaporation has been neglected.}
\end{figure}

\begin{figure}
\centering
\epsfxsize=8cm
\epsfbox{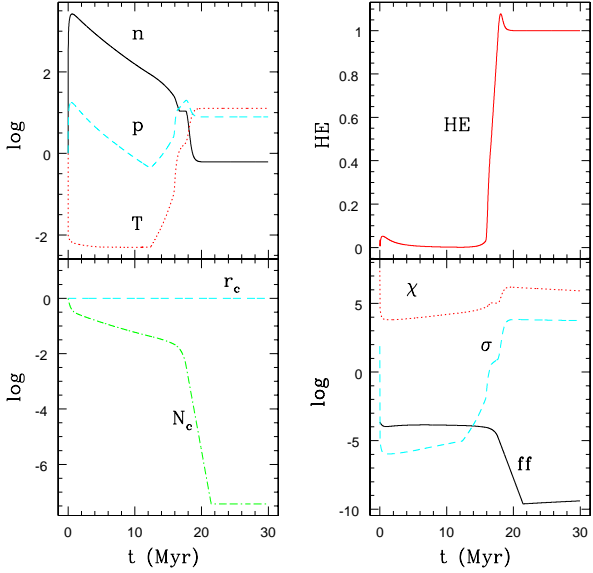}
\caption{Model 5. The same as in Figure 7 (Model 1), except that the initial
single mass for the clouds has been replaced by a mass distribution function
with spectral index $\alpha = 1.5$ (see Ap. A).}
\end{figure}

\begin{figure}
\centering
\epsfxsize=8cm
\epsfbox{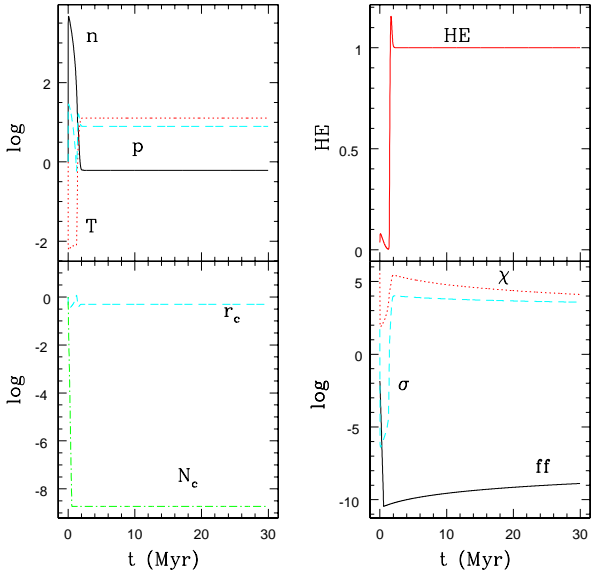}
\caption{Model 6. The same as in Figure 7 (Model 1), except that only
the first generation of clouds is present in the SB ambient medium, and no
further clouds are allowed to form.}
\end{figure}

In summary, the results above show that, provided that a continuous
formation of clouds due to deposited material on the SNR shells occurs in the
SB, then the common assumption of a HE value
close to 1 is reasonable only after many Myr, when a large quantity of
matter has
been blown-out and the ambient gas density has dropped to values of $\leq
10^{-1}$ cm$^{-3}$ (Note that in Figs. 7 to 13, the ambient densities $n$
have been normalized to their initial values which are given in Table 2.).

\section{Discussion and Conclusions}
The high stellar density in a SB galaxy allows for SNRs interactions that
very quickly may fill all the ambient, forming a superbubble containing hot
gas with temperatures T=$10^4$-$10^6$ K, and densities $n$=0.1-0.01
${\rm cm^{-3}}$.
Shells interactions increase the process of clouds formation with typical
temperatures in the range of 100-$10^4$ K, depending on whether they are
photoionizated or not.
Under these conditions, hot gas may eventually escape from the galaxy in an
outflow that changes the chemical and dynamical characteristics of the galaxy
itself.
It is clear that single SN explosions have a typical efficiency in the
energization of the ISM of normal galaxies of few per cent, but a debate
exists about the SN heating efficiency in SB galaxies.
To solve this problem, we have developed a simple semi-analytical model that
analyses the most important physical processes in a three-component system
formed by hot gas, SNRs and cold (or warm) clouds. Under this condition, the
evolution of the three-phase medium after the development of an instantaneous
SB has been examined to compute the SN heating efficiency (HE).
The study has also been accompanied by 3-D numerical experiments involving the
interaction between SNRs and clouds in a SB environment.
We have found (see $\S$ 5) that HE remains very low as long as the
mass-loss time scale of the clouds ($t_{mx}$) remains smaller than the time
scale for a superbubble formation ($t_{int}$).
For an environment with continuous formation of clouds, this occurs
during
the first 16 Myrs for an initial total gas mass in the clouds
of 8 $\times 10^6 M_{\odot}$, after which HE rapidly increases to unity.
This result can be explained by the fact that the efficient ablation of the
clouds during the first $\sim$ 16 Myrs increases the density and the radiative
cooling of the gas of the ambient
medium, therefore preventing a temperature enhancement (HE $\sim$ 1) that
would lead to gas expansion with the formation of a hot and rarefied
superbubble and a sudden expulsion of the gas through a galactic wind.

This result is quite insensitive to the initial conditions of the hot diffuse
ambient ISM of the SB, nonetheless the semi-analytical calculations show that
the SN heating efficiency can be affected by parameters such as the total gas
mass in the clouds.
We find that an increase in this parameter results low HE values
for longer times. In fact, an increase in this parameter by a factor 10 with
respect to the value above makes HE $<$ 1 over the entire SB lifetime.

We have also found that under an extreme condition in which no further
clouds form after
the first generation of clouds, then this initial population alone is unable
to
prevent a faster raise of the ambient gas temperature and therefore, of ,
the HE value to one. However, such situation could be expected to occur only
in
environments which are initially already  completely ionized by the UV photons
from massive stars, having no cold clouds or shells.

In summary, we can conclude that the SN efficiency has a time-dependent
trend which is sensitive to the initial conditions of the SB system.
Therefore, it is not possible to assume a single value for HE over the entire
SB lifetime without considering the physical interactions that can change the
environmental characteristics.
If all the  gas that is swept by the SNR shells goes into clouds
formation, providing a continuous generation of clouds, then
HE remains small up 16 Mrys. If, on the other hand,
the material that is swept by the SNRs does not lead to new generations
of clouds, then the gas forms a superbubble in few Myrs.
16 Myr is, therefore, an upper limit  for the time it takes for HE to increase
to unity.

Our model allows one to obtain more realistic HE values depending on the
evolution stage of a SB galaxy. In particular, it provides a natural
explanation for the low values invoked in some models to describe the chemical
evolution of dwarf galaxies (Carigi et al. 1995; Bradamante Matteucci \&
D'Ercole, 1998;
This is the case, for example, of the SB galaxy SBIZw18 (Recchi,
Matteucci \& D'Ercole 2001), for which a value of $HE \sim$
0.03 is required to explain observations and is in good agreement with the
results of this work which predicts $HE \sim$ 0.01 to 0.1 in the first
$\leq$ 16 Myr of the SB lifetime (Figs. 7 - 12).

We have conducted our study considering regions which are small compared to
the total size of the host galaxy.
There must be other relevant physical phenomena outside of this region that
can also influence its evolution and the determination of HE, as for example,
the rate at which material is accreted from the neighborhood into the SB
region.
In fact, during some period of time, SBs may be effectively fed by gas from
outside.
The 2D hydrodynamic models of galactic winds in SB galaxies found in
the literature often produce a superbubble that exceeds the dimensions of the
narrow cones detected in the nuclear SB, but Tenorio-Tagle \&
Mu\~noz-Tu\~n\'on (1998) have shown that, if the gas of the galactic disk
is accreted onto the SB region at a rate of a few $\msun$ yr$^{-1}$,
the funnels may present small opening angles, in agreement with the
observations. This infall is thought to be necessary to
account for the accumulation of the matter detected in these sources
(Sanders \& Mirabel 1996) and the large rates of star formation
(Larson 1987 et al., Suchkov et al. 1994).
In the present study, although these effects of matter accretion from the
neighborhood have not been taken into account, we expect that their presence
will keep the HE value smaller than unity for even larger times, since
they must cause an increase in the formation rate of clouds.

Recently, IR, radio, continuum and UV observations of SBs
have revealed the presence of very compact structures with typical sizes of
few parsecs and mass in the range of $10^4-10^5 M_{\odot}$,
that can be associated with star clusters and super star clusters (see,
e.g., Ho 1997; Johnson et al. 2001; Gorjian, Turner \& Back 2001). This is
somewhat different from the general assumption often employed in the modeling
of SBs and also adopted in the present work, that the star formation rate is
diluted over all the SB volume. However, even if the bursts are not
homogeneously distributed, we expect that the results here obtained will not
be much affected.
Since each star cluster must
drive a bubble with morphological characteristics similar to a SNR, when these
bubbles interact with each other the global evolution within the SB will be
similar to the one analyzed in the present work.

Another relevant aspect that deserves attention is the determination of the
conditions of the galactic wind that develops when HE $\sim 1$.
In this work we have assumed a constant rate for the SN energy injection over
the SB lifetime. This assumption is consistent with previous modeling of SB
systems (see, e.g., Leitherer et al. 1999).
However, if we had taken into account another possible evolution scenario,
with a decreasing SN energy injection rate with time, then in this case, the
reduced available SN injection power in the late stages could be not
sufficient
to drive the galactic wind that, according to our model, should evolve only
in the second half of the SB lifetime.
This may offer a potential explanation of why some SB galaxies produce
winds and others do not.
The computation of the total quantity of mass that is transported by the
galactic wind (Suchkov et al. 1994) is also dependent of the physical history
of the expelled clouds.
Although at a different rate, they will be also ablated by photoevaporation,
drag and thermal evaporation, contributing to the growth of the matter that
forms the outflow itself.

A more particular attention should be given to magnetic fields, which
can be important in reducing the ablation processes and which can influence
the escape velocity of the hot gas and the cold clouds.
Although in the present work they have been taken into account in the
 computation of the total internal pressure of the clouds (eq. 6) and also
explicitly in the evaluation of the their dominant ablation
process (the photoevaporation), they have not been included in the
evaluation of the pressure of the ambient gas and the SNRs shells,
or in the numerical simulations of SNR-clouds interactions. Also,
to be remarked is the fact that the adopted value, of $\sim$ few
$\mu G$ within the clouds, is somewhat uncertain since, to our
knowledge, there is no direct magnetic field measurements in SBs.
Nonetheless, it is consistent with the values often required for
pressure equilibrium of the clouds with the hot ambient medium in
our Galaxy, and also with the lower limit strengths necessary to
collimate the galactic winds in SBs (de Gouveia Dal Pino \& Medina
Tanco 1999). In the case of the SNRs, while the presence of
magnetic fields in the shells could, on one side, reduce their
compressibility and thus decrease the volume of hot gas within the
remnant (see, e.g., Slavin \& Cox 1993, McKee \& Zweibel 1995), on
the other side, it could constrain the thermal conduction and thus
inhibit the destruction of this hot gas component (e.g., McKee
1995). The reduction of compression by the magnetic field could
also inhibit the process of shell fragmentation during
interactions with other SNRs and, consequently, the formation of
clouds by this process (that was discussed in $\S$ 3.1). The
relative importance of  these competing effects may be quantified
after determination, from observations, of the magnitude and
geometry of the magnetic fields in SB environments.

Finally, we should notice that studies that search for a potential connection
between SBs and AGN activity, that suggest that SBs at the nuclear regions of
active galaxies could eventually trigger the development of a supermassive
black hole (BH) in the center, may be favored by the results of the present
analysis as these predict a delay in the expulsion of gas from the SB in form
of a galactic wind, for $\sim$ 16 Myrs.
The gas retained in the system may lead to new bursts of star formation and
increasing deposition of material in the center of the SB that, in turn, may
eventually form a BH.

\begin{acknowledgements}
C.M. and E.M.G.D.P acknowledge financial support from the Brazilian Agencies
FAPESP and CNPq.
The authors are deeply indebted to A. D'Ercole for his giving the original
idea for the work and for his useful and uncountable comments and suggestions
during its elaboration. The authors also acknowledge useful suggestions
of the referee that have helped to improve the paper.

\end{acknowledgements}

\appendix
\section{}

The cumulative cloud mass-loss rate in a SB may be computed taking into
account the spectrum of the clouds number $N (m)\propto \mc^{-\alpha}$
between an upper ($m_{\rm c,sup}$) and a lower ($m_{\rm c,inf}$) limit,
instead of considering only a single cloud mass.
Observations of high star formation regions indicate a minimum mass of
$10^{-4} M_{\odot}$ and a maximum mass of $\sim 10^3 M_{\odot}$
(Kramer et al. 1998).
This is consistent with the theoretical maximum stable mass of a magnetically
dominated cloud against gravitational collapse (Bertoldi \& McKee 1990).
We thus assume $m_{\rm c,sup}=10^3$ $\msun$ and
$X=m_{\rm c,inf}/m_{\rm c,sup}= 10^{-7}$.
Also, from the studies of Zinnecker, McCaughrean \& Wilking (1993)
and Blitz (1993), we take $\alpha=1.5$.
A plot for this cloud mass distribution is shown in Figure A1.

\begin{figure}
\centering
\epsfxsize=8cm
\epsfbox{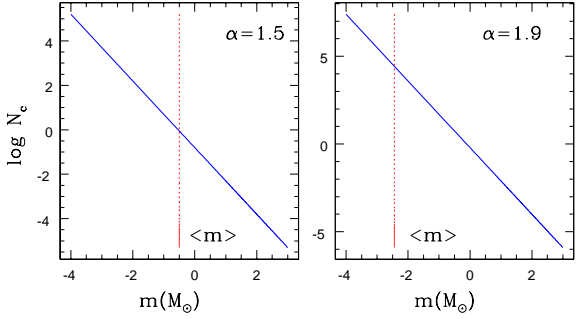}
\caption{Initial number of clouds for a distribution of the clouds number in
a range of $10^{-4} - 10^3 M_{\odot}$,
and the value of the average mass $<m>$ of the distribution $N_c \propto
m_c^{-\alpha}$, with $\alpha$ = 1.5 (left), and  $\alpha$ = 1.9 (right).}
\end{figure}

Let $f(\mc)=\theta \mc^{\gamma}$ be the mass-loss-rate of a single cloud,
where the constants $\theta$ and $\gamma$ depend on the specific process
responsible for the cloud mass-loss studied in $\S$ 3 (i.e. thermal
evaporation, drag or photoevaporation).
The cumulative mass-loss rate must
be computed taking into account the mass spectrum of the clouds above
$d N_c(m_c)/dm \propto \mc^{-\alpha}$. The total mass-loss rate per process
is thus given by
\[
 \dot M=\!\!\int_{m_{\rm c,inf}}^{m_{\rm c,sup}}\!\! N_c(\mc)f(\mc)\>
{\rm {d}}\mc,
\]
\noindent
and can be written as
\[
\dot M=\nc f(m_{\rm c,sup}) g(\alpha, \gamma, X)
\]
\noindent
where $N_c$ is the total number of clouds inside the SB region and
\[
 g(\alpha, \gamma, X)={1-\alpha\over 1+\gamma -\alpha}{1-X^{1+\gamma -\alpha}
\over 1-X^{1-\alpha}}.
\]
Taking into account Equations (7), (12) and (16) for $\dot M_{UV}$,
$\dot M_{d}$ and $\dot M_{ev}$ respectively, we obtain $\gamma$ = 12/21, 16/21
and 8/21 for cloud photoevaporation, drag and thermal evaporation  processes,
respectively.
We find that g($\alpha, \gamma, X$) is rather sensitive to the parameters
involved.
Figure 14 shows $g(\alpha, \gamma, X)$ as a
function of $X$, for $\alpha=1.5$, $\alpha=1.9$, and the three values of
$\gamma$.

For $\flx=0.225$ (see eq. 6), we
obtain $\dot M_{\rm UV}=2.9\times 10^{-6}M_g$ $\msun$ yr$^{-1}$,
$\dot M_{\rm ev}=6.17\times 10^{-12}T_6^{5/2}M_g$ $\msun$ yr$^{-1}$,
and $\dot M_{\rm d}=1.32\times 10^{-12}n \chi^{1/2}v_6
M_g$ $\msun$ yr$^{-1}$. As demonstrated in Section 3,
photoevaporation is by far the most important mechanism for the cloud
mass-loss rate during the first Myrs of the SB life, even considering a cloud
mass distribution, and the results do not differ much from those when one
takes a single initial mass (see section 5).

\begin{figure}
\begin{center}
\begin{tabular}{cc}
\epsfxsize=4cm
\epsfbox{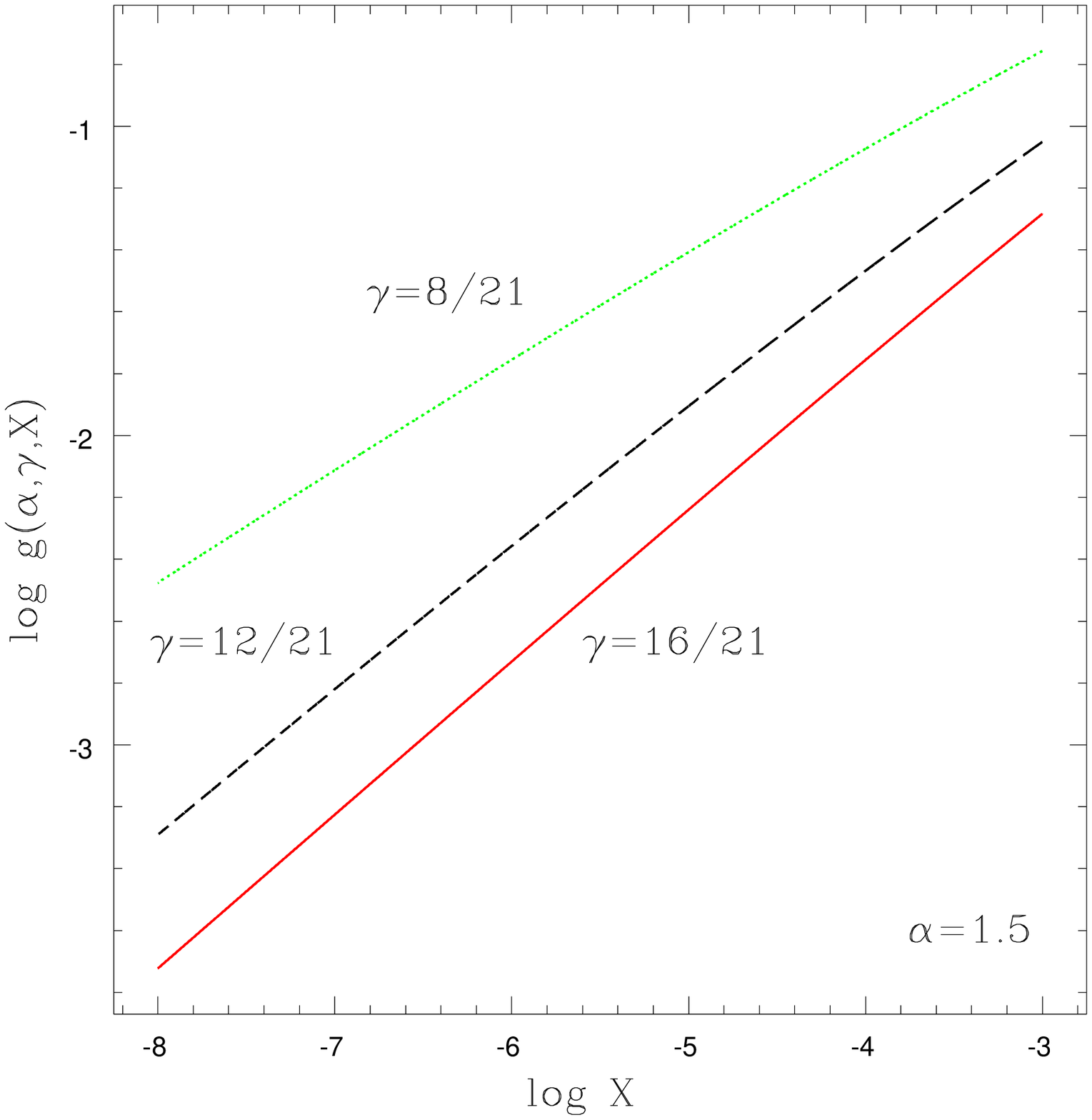} &
\epsfxsize=4cm
\epsfbox{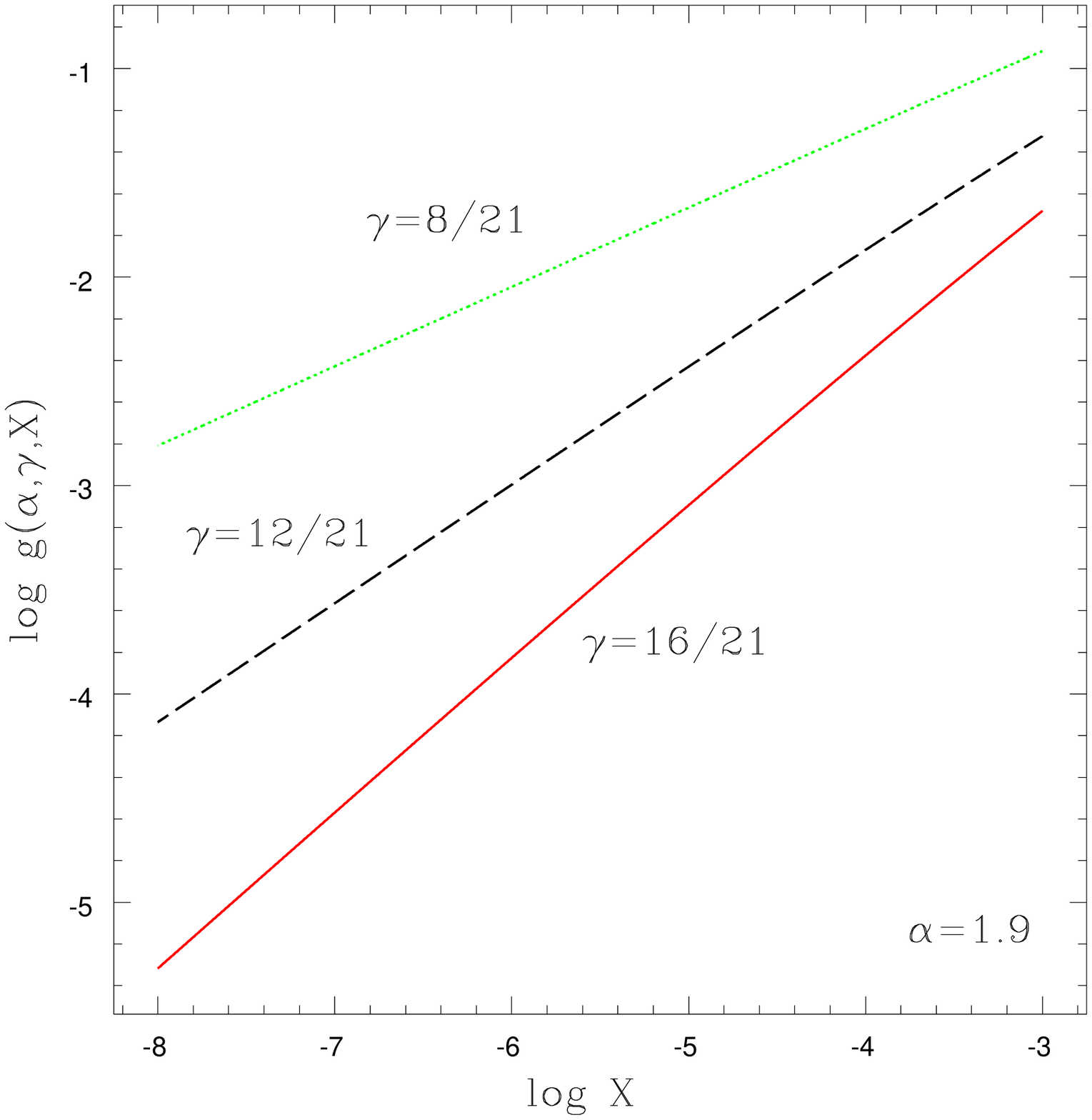}
\\
\end{tabular}
\caption{The function $g(\alpha, \gamma, X)$ for $\alpha =1.5$ (left
panel) and $\alpha =1.9$ (right panel) for three different values of
$\gamma$.}
\end{center}
\end{figure}

\section{}

We present in this Appendix the explicit expressions for the constants
$C_{\rm {n}}$, with n = 1 to 10, that appear in equations 25, 26, 27, 28
in $\S$ 5.

\begin{equation}
C_1 = {{4.64 \times 10^{14} \ V_{SB} m_{\rm {H}} \mu} \over
{m_c}}
\end{equation}
\noindent

\begin{equation}
C_2 = {\delta \over {3.15 \times 10^7}}
\end{equation}
\noindent
where $\delta$ is the fraction of gas that is accumulated in the SNR shell.

\begin{equation}
C_3 = {1 \over {3.15 \times 10^7 \ m_c}}
\end{equation}
\noindent

\begin{equation}
C_4 = {1 \over {4.6 \times 10^{29} \ V_{SB}}}
\end{equation}
\noindent
where $V_{SB}$ is the SB volume

\begin{equation}
C_5 = {S_{SB} \over {V_{SB}  \ (3.08 \times 10^{18})}}
\end{equation}
\noindent
where $S_{SB}$ is the spherical external surface of the SB in pc.

\begin{equation}
C_6 = {{1.1 \times 10^{-24}} \over {V_{SB} m_{\rm {H}} \mu}}
\end{equation}
\noindent
which gives the increase in the density due to the mass-loss rate
from the stellar winds and the SNe explosions.

\begin{equation}
C_7 = {{7.24 \times 10^{-64}} \over V_{SB}}
\end{equation}

\begin{equation}
C_8 = {{4.56 \times 10^{-23}} \over {V_{SB}}}
\end{equation}

\begin{equation}
C_9 = {2 \over 3}
\end{equation}

\begin{equation}
C_{10} = {{2 S_{SB}} \over V_{SB}}
\end{equation}

All the dimensional quantities of the SB ($V_{SB}, S_{SB}, R_{SB}$) are
expressed in pc. The mass-loss rate of the clouds is in $M_{\odot}$/yr,
the time scales are in yrs, and we $m_c$, that is assumed constant, in
M$_{\odot}$.

{}

\label{lastpage}
\end{document}